\documentclass[
aps,%
12pt,%
final,%
notitlepage,%
oneside,%
onecolumn,%
nobibnotes,%
nofootinbib,%
superscriptaddress,%
noshowpacs,%
centertags]%
{revtex4}

\usepackage{epsfig}

\begin{document}

\pagenumbering{arabic}

\title{
Optimal values of rovibronic energy levels for triplet electronic states
of molecular deuterium}

\author{B.~P.~Lavrov}
\email{lavrov@pobox.spbu.ru}
\author{I.~S.~Umrikhin}
\affiliation{
Faculty of Physics, St.-Petersburg State University, \\
St.-Petersburg, 198504, Russia}

\begin{abstract}

Optimal set of 1050 rovibronic energy levels for 35 triplet 
electronic states of $D_2$ has been
obtained by means of a statistical analysis of all available
wavenumbers of triplet-triplet rovibronic transitions studied in emission,
absorption, laser and anticrossing spectroscopic experiments of
various authors. We used a new method of the analysis (Lavrov,
Ryazanov, JETP Letters, 2005), which does not need any \it
a~priory \rm assumptions concerning the molecular structure being
based on only two fundamental principles: Rydberg-Ritz and maximum likelihood. 
The method provides the opportunity to obtain the RMS
estimates for uncertainties of the experimental wavenumbers independent
from those presented in original papers. 234 from 3822
published wavenumber values were found to be 
spurious, while the remaining set of the data may be divided into
20 subsets (samples) of uniformly precise data having close 
to normal distributions of random errors within the samples.
New experimental wavenumber values of 125 questionable lines 
were obtained in the present work. 
Optimal values of the rovibronic 
levels were obtained from the experimental 
data set consisting of 3713 wavenumber values (3588 old and 125 new).
The unknown shift between levels of ortho- and para-
deuterium was found by least squares analysis of the
$a^3\Sigma_g^+$, $v = 0$, $N = 0 \div 18$ rovibronic levels with 
odd and even
values of $N$. All the energy levels were obtained relative to the
lowest vibro-rotational level ($v = 0$, $N = 0$) of 
the $a^3\Sigma_g^+$ electronic state,  
and presented in tabular form together with the standard deviations 
of the empirical determination. New energy level values 
differ significantly from those available in literature.
\end{abstract}

\maketitle

\section{Introduction}

The present work is devoted to the studies of the wavenumbers of the
triplet-triplet radiative electronic-vibro-rotational (rovibronic) 
transitions and empirical determination of the optimal set of
the triplet rovibronic energy levels of the $D_2$ molecule. 

Studies of spectra and structure of molecular deuterium represent 
not only understandable general interest (an isotopomer of simplest 
neutral diatomic molecule), but have also direct practical value in 
connection with wide use of $D_2$ in physical experiments and in 
various technical applications: from gas-discharge sources of ultraviolet 
radiation \cite{GLT1982} up to tokamak-reactors \cite{PBS2005}.

The spectrum of the $D_2$ molecule is caused by both singlet-singlet 
and triplet-triplet rovibronic transitions. The intercombination 
lines were not observed yet. The most interesting resonance singlet band 
systems are located in vacuum ultraviolet (VUV), while the triplet 
transitions are responsible for major part of light emission of 
ionized gases and plasma in near infrared, visible and near ultraviolet. 
They are often used for spectroscopic diagnostics of non-equilibrium plasmas 
\cite{GLT1982, PBS2005, LKOR1997, LMKR1999, RDKL2001, LPR2006}. 
Therefore, the triplet rovibronic levels and optical transitions between 
them were chosen as the object of the present research.

The energy level values evaluated from experimental wavenumbers are often 
called \it observed \rm energy levels irrespective of a method used for 
the data processing. Such rather disputable terminology implicitly assumes 
that all spurious experimental results and systematic errors are absent, 
and random errors are negligibly small (Otherwise, several different sets 
of the observed energy level values may be obtained from the same set of experimental 
wavenumbers by applying different methods of the data processing.). Therefore, 
we prefer to use the term \it empirical \rm level values 
\footnote{Compare with the term \it experimental \rm level values in \cite{Wetal2006}}
(for any values obtained from experimental data) and \it optimal \rm level 
values (for the values obtained by pure statistical approach 
\cite{LavrovRiazanovJetf, LavrovRiazanovOiS}), leaving 
the word \it observed \rm for experimental values of the wavenumbers 
which are unambiguously linked with the wavelengths --- real observables 
in spectroscopy. (It should be also mentioned that in molecular 
spectroscopy energy levels in cm$^{-1}$ are often called molecular terms 
(see e.g. \cite{Kovacs1969}). Absolute values of both levels and terms
are always positive and calculated relative to the lowest 
vibro-rotational level of the ground electronic state considered as a 
zero level. That is in contrast to original meanings of terms and 
energies adopted in atomic spectroscopy.)

Generally speaking, the best set of empirical values of energy may be 
named a set of optimum values (optimal for available experimental data)
only if these values do not depend on a procedure of their obtaining, 
and are entirely defined only by amount and quality of available 
experimental data. In such favourable case the energy 
level values may be considered as 
secondary experimental data.

It is well known that diatomic hydrogen, being the simplest neutral
molecule, has a most sophisticated emission spectrum. The hydrogen band
spectrum, caused by spontaneous emission due to
rovibronic transitions, 
does not show a visible, easily recognizable band structure, but
has the appearance of a multiline atomic spectra. The peculiarity of
molecular hydrogen and its isotopic species --- abnormally small
nuclear masses --- leads to high values of vibrational and
rotational constants and large separation between vibrational and
rotational levels of various excited electronic states. As a
result, various rovibronic spectral lines belonging to different
branches, bands and band systems are located in the same 
spectral regions, leading to the overlap of various band
systems, bands and branches, as well as the mixing of 
rovibronic spectral
lines having different origins. The small nuclear masses 
stimulate a breakdown of the Born-Oppenheimer approximation due
to electronic-vibrational and electronic-rotational perturbations
having both regular and irregular character; this combination 
seriously
complicates the interpretation of the spectra of hydrogen isotopomers
and the unambiguous identification of rovibronic spectral lines.
Symmetry rules for permutation of identical nuclei in homonuclear
isotopomers ($H_2$, $D_2$ and $T_2$) cause the known effect of the 
intensity alternation of neighbouring lines within the rotational 
structure of bands
due to the alternation in degeneracy of successive rotational
levels with odd and even values of rotational quantum number
(e.g. 1:2 in the case of $D_2$). This effect also masks the
visible structure of branches resulting in serious
additional difficulties for identification of rovibronic 
spectral lines. 

Thus, most of the lines in the optical spectra of hydrogen
isotopomers have not yet been assigned in spite of tremendous
efforts by spectroscopists over the previous century 
\cite{Richardson, Dieke1958, Dieke1972, BredohlHerzberg1973, Huber, Crosswhite, RossJungenMatzkin2001}. 
As an example, in the latest
compilation  of experimental data for molecular deuterium $D_2$ 
\cite{Crosswhite}, the working list of 27488 recorded lines
contains only 8243 assignments. These assignments were obtained by
traditional methods of analysis using wavenumber combination
differences (method of common differences) 
and Dunham series expansions \cite{Richardson, Dieke1972}, sometimes 
together with comparison of molecular constants obtained for
different isotopic species \cite{Dieke1958}. Later on, the traditional 
methods were supplemented by taking into account the line intensities
\cite{LO1978} and the results of \it ab~initio \rm and semi-empiric 
calculations \cite{Crosswhite, RossJungenMatzkin2001}. 

Recently in \cite{RLTBjcp2006, RLTBjcp2007}, 
the identification of rovibronic lines of three singlet 
VUV band systems of $D_2$ has been obtained by common use of 
semi-empiric calculations (of the levels and the transition probabilities)
together with the program IDEN \cite{Azarov1991, Azarov1993} developed for 
visual identification of complex atomic spectra.
Although the approach seems to be a rather powerful and
prospective tool for finding new assignments in the spectra of 
hydrogen isotopomers, the final results of \cite{RLTBjcp2006, RLTBjcp2007} 
can not be considered as optimal values of the rovibronic energy levels for
the $D^1\Pi_u$, $D'^1\Pi_u$, and $B'^1\Sigma_u$ electronic states studied in
\cite{RLTBjcp2006, RLTBjcp2007}.
They were simply calculated as the average of values derived from 
different observed line wavenumbers added to the corresponding lower 
vibro-rotational levels of the $X^1\Sigma_g^+$ ground electronic 
state from \cite{BredohlHerzberg1973}, assumed to be 
known with negligible uncertainty. 
Actually, those lower levels were obtained from wavelengths of two
other VUV band systems
($B^1\Sigma_u^+ - X^1\Sigma_g^+$, $C^1\Pi_u^- - X^1\Sigma_g^+$) 
by two independent methods of the analysis which gave
noticeably different results (compare columns 2 and 4 of table 4 from
\cite{BredohlHerzberg1973}). It is evident that common use of experimental 
data for five singlet band systems mentioned above may give another set of
rovibronic levels for $X^1\Sigma_g^+$ ground electronic state.
Then the level values for the exited states will be changed somehow.
In our opinion, the assignments
found in \cite{RLTBjcp2006, RLTBjcp2007} 
also should be confirmed by the appropriate 
statistical analysis of all available in literature experimental data 
on the wavenumbers for all other singlet band systems of $D_2$. For 
that purpose it is possible to use the approach proposed and realized 
in our papers \cite{LavrovRiazanovJetf, LavrovRiazanovOiS, BHterms, ALMU2008}.

Currently, available information concerning triplet rovibronic energy 
level values of $D_2$ molecule exists in the form of the list of molecular
constants for Dunham series expansions in \cite{Huber, Nist},
and tables of rovibronic levels obtained in \cite{Crosswhite}.

In the case of the hydrogen molecule the Dunham coefficients are known
to provide a very poor description of the rovibronic energy level 
values \cite{LPU1986}. Moreover any set of molecular constants depends 
not only on molecular properties but also on the theoretical model 
used in the analysis of the spectrum (the type and length of the 
series expansion, form of a model Hamiltonian, etc.), as well as on amount of 
measured wavenumbers and even on the distribution of the spectral 
lines over the vibrational and rotational levels (see e.g.
\cite{LavrovRiazanovJetf, LavrovRiazanovOiS, LPU1986, Albriton}). 
Therefore, the molecular constants are rather conditional and the 
procedures used for their determination are subjective in character. 
In practical use of reference data on molecular constants, the 
question arises as to how many rovibronic
levels of a particular type can be described by a
given set of constants with a required accuracy. In the
best case, this question can be answered only by the
researcher who obtained the given set of molecular
constants.
However, series expansions and molecular constants may be useful 
both in a process of classifying rovibronic lines in unknown spectra 
and in calculations which do not pretend for a high accuracy.

The data reported in \cite{Crosswhite}, in general, give a rather 
good description of the $D_2$ spectrum, but 
they are also not free from criticism. The method of the analysis 
used in \cite{Crosswhite} is based on the common use of the 
combination differences, some selected wavenumbers for certain 
transitions and one by one multistage 
treating of separate branches, bands and band systems. 
The sequence of the steps chosen in \cite{Crosswhite} is not 
the only possible analytical arrangement. Therefore, the data thus 
obtained can not be considered as an optimal set of levels providing 
the best description of all observed wavenumber values. 
On our opinion the reported in \cite{Crosswhite} uncertainty 
$\sigma = 0.05$ cm$^{-1}$ for all levels and all wavenumbers of
lines is the overestimation caused by the randomization of various 
deviations when working with many thousands of levels and lines.
It should also be 
mentioned that after publication of \cite{Crosswhite}, new 
experimental data on the wavenumbers appeared \cite{DabrHerz, Davies}.

Recently, a new method of a pure statistical analysis of experimental 
data on the rovibronic transition wavenumbers for empirical determination 
of the optimal energy-level values has been proposed 
\cite{LavrovRiazanovJetf} and successfully applied for the derivation of 
rovibronic level values of the singlet states of the $BH$ \cite{BHterms} 
and the triplet states of the $H_2$  \cite{ALMU2008} molecules.
The method is based on only two
fundamental principles: Rydberg-Ritz and maximum likelihood. 
This approach differs from known techniques in several aspects: 
1) does not need any assumptions concerning an internal
structure of a molecule; 
2) does not employ intermediate parameters, such as molecular 
constants in the traditional methods; 
3) a one-stage optimization procedure can be used for all available 
experimental data obtained for various band systems, by various 
methods and authors, and in various works (Simultaneous adjustment 
of all levels eliminates the possibility of accumulative errors 
in traditional methods, caused by multi-step treating of various bands
one by one, as well as by transfer of any occasional error for one 
level, appeared on one step, for all higher levels 
obtained during further steps);
4) provides the opportunity of a rational censoring of the experimental
data in an interactive mode (thus allowing the user the option to 
eliminate obvious errors (outliers) and misprints, to revise incorrect
line assignments, and to
compare various sets of experimental data for mutual consistency);
5) gives an opportunity of independent estimation
of experimental uncertainties by analyzing the shapes of 
error distributions 
within various samples of the experimental data
(It means that our statistical estimates of the 
root-mean-square (RMS) experimental errors are totally independent 
on the estimates reported in original experimental papers); 
6) provides an optimal set of rovibronic level values as well as the
uncertainties of their determination (standard deviations SD and the 
covariance matrix) caused only by the amount and quality of existing
experimental data \cite{LavrovRiazanovOiS}.

A necessary precondition for an application of the method 
\cite{LavrovRiazanovJetf} is an existence of more or less dependable 
assignments for majority of rovibronic spectral lines under the study. 
Therefore, the new method should be considered as complimentary for the 
traditional and new methods of identifying unknown spectra. However, 
since the new, pure statistical method is objective, it is better to 
consider any identification of a rovibronic spectrum as preliminary until 
it is not corroborated by that, the use of certain assignments in the
framework of this method does not lead to contradictions with the 
Rydberg-Ritz principle or with normal distribution of random experimental
errors.

The goal of the present paper is to report the results of applying
the new method \cite{LavrovRiazanovJetf} for statistical analysis
of the rovibronic spectral line wavenumbers of triplet band systems
and determining the optimal set of rovibronic energy levels for
all known 35 triplet electronic states of molecular deuterium:
$a^3\Sigma_g^+$, $c^3\Pi_u^+$, $c^3\Pi_u^-$, $d^3\Pi_u^+$,
$d^3\Pi_u^-$, $e^3\Sigma_u^+$, $f^3\Sigma_u^+$, $g^3\Sigma_g^+$,
$h^3\Sigma_u^+$, $i^3\Pi_g^+$, $i^3\Pi_g^-$, $j^3\Delta_g^+$,
$j^3\Delta_g^-$, $k^3\Pi_u^+$, $k^3\Pi_u^-$, $n^3\Pi_u^+$,
$n^3\Pi_u^-$, $p^3\Sigma_g^+$, $q^3\Sigma_g^+$, $r^3\Pi_g^+$,
$r^3\Pi_g^-$, $s^3\Delta_g^+$, $s^3\Delta_g^-$, $u^3\Pi_u^+$,
$u^3\Pi_u^-$, $(7p)^3\Pi_u^+$, $(7p)^3\Pi_u^-$, $(8p)^3\Pi_u^+$,
$(8p)^3\Pi_u^-$, $(9p)^3\Pi_u^+$, $(9p)^3\Pi_u^-$,
$(6d)^3\Sigma_g^+$, $(7d)^3\Sigma_g^+$, $(8d)^3\Sigma_g^+$ and
$(9d)^3\Sigma_g^+$.

\section{Statistical approach to empirical determination of 
optimal values of the rovibronic energy levels}

In principle an analysis of experimental data on wavenumbers of 
spectral lines may be considered as consisting from two separate 
parts: 
1) an identification (classification, assignment) of spectral 
lines (an establishment of a correspondence between observed 
spectral lines and pairs of the initial and final levels 
responsible for occurrence of these lines in a spectrum), and 
2) the determination of energy level values from measured 
wavenumbers (or some other magnitudes related to the levels, 
like molecular constants \cite{LPU1986}, or potential curves 
\cite{DLPU1987}). The aim of the first stage of an analysis is 
to find most likely identification and to prove correctness 
of the assignments. The goal of the second is to obtain the 
best possible values of the energy levels as well as the 
uncertainties of their empirical determination. In spite of 
the difference in objectives these two parts of an analysis 
are closely connected because quality of final results in both 
cases is determined by their ability to reproduce observed spectra. 
Therefore, quite often the parts are not distinguished in original 
papers.

The main part of the present work represents the second stage of the 
analysis of the triplet-triplet spectrum of $D_2$ molecule, based on 
the results of the first stage reported in 
\cite{Crosswhite, DabrHerz, Davies, DiekeBlue, Dieke, GloersenDieke, DiekePorto,  
FreundMiller}.
\footnote{In some cases discussed below, we have had to do the job of 
the first stage when it was necessary to change earlier assignments 
and to obtain new experimental values of the wavenumbers of some 
questionable lines.}
Therefore in the further consideration it is supposed, that more or 
less reliable identification is already established for the majority 
of lines. Then not numerous cases of wrong assignments may be revealed 
as outliers and corrected during the statistical analysis (see below).

All spectroscopic methods of empirical determination of energy levels of 
atoms and molecules are based on the Rydberg-Ritz combination principle 
corresponding to 
the Bohr frequency rule in quantum mechanics. For rovibronic transitions of 
diatomic molecules between electronic states corresponding to Hund's case "b" 
for angular momenta coupling (negligible multiplet splitting) it may be written as:

\begin{equation}
\nu^{n'v'N'}_{n''v''N''} = E_{n'v'N'} - E_{n''v''N''}, \label{ridbergritz}
\end{equation}
where $n$ indicates an electronic state, and $v$ --- the vibrational quantum number.
In case of diatomic hydrogen isotopomers the multiplet splitting 
and hyperfine structure of levels and lines are negligibly small and 
usually unresolved in experiments, therefore in the present work 
rotational levels are characterized by the quantum number $N$ of the 
total angular momentum of a molecule excluding spins of electrons and nuclei.
Upper and lower rovibronic levels are marked by single and double primes, 
respectively. The $\nu^{n'v'N'}_{n''v''N''}$ is the wavenumber 
(recalculated for vacuum conditions)
of the n', v', N' -- n'', v'', N'' rovibronic transition, and $E_{nvN}$ are 
corresponding energy level values.

One may see that the experimentally measured wavenumbers of the spectral 
lines are related only to the differences between pairs of rovibronic energy 
levels. So the values of the levels cannot be directly obtained by means of 
the Rydberg-Ritz principle only. Therefore, traditional methods of empirical
determining the rovibronic energy levels require the introduction of some 
additional (with respect to the Rydberg-Ritz combination principle) assumptions
regarding properties of molecules and, hence, are semiempirical and allow some 
subjectivity of a researcher (see details in \cite{LavrovRiazanovOiS}).

The classification of the levels and lines by certain sets of quantum numbers 
($n$, $v$, $N$) is important for the assignment of the wavenumbers to the 
rovibronic levels between which the transitions occur and, in particular, 
for the application of the selection rules. When this assignment is carried 
out, the specific designations of the levels and lines are not important any more. 
The rovibronic levels may be denoted by natural numbers in an arbitrary order. 
The notation $E_{nvN}$ may be changed to a more compact notation $E_{i}$, and 
the wavenumbers are designated by pairs of indices corresponding to the initial 
and final levels. Then the expression (\ref{ridbergritz}) becomes:

\begin{equation}
\nu_{ij} = E_i - E_j. \label{ridbergritz2}
\end{equation}

Suppose that the set of available experimental data consists of $n_{\nu}$ 
wavenumber values obtained for $n_T$ transitions caused by the combinations 
of $n_E$ energy levels. $n_{\nu} > n_T > n_E$ because an amount of the 
transitions allowed by selection rules usually considerably exceeds quantity 
of combining levels. Furthermore, experimental data for the same transitions 
may be obtained and reported in various publications. 
One may insert all experimental wavenumbers into left hand side of the 
(\ref{ridbergritz}) one by one, and consider the energy levels 
as adjustable parameters to be obtained from the experimental data. Then 
(\ref{ridbergritz}) turns into a system of $n_{\nu}$ equations 
with $n_E$ unknown quantities $E_{i}$. This system of equations is 
overdetermined and, hence, is inconsistent because the experimental 
data always involve measurement errors.
	
Straightforward general solution for solving such problems is well-known 
in mathematical statistics. That is the least squares method based on the 
assumption of finite second moment of the distribution function for random 
errors \cite{Hudson}. In our case it consists in the minimization of the 
weighted mean-square deviation between observed $\nu_{ij}^{expt}$ and 
calculated (as differences of adjustable energy levels $E_i$, $E_j$) 
values of rovibronic line wavenumbers, or the sum

\begin{equation}
r^2=\sum_{\nu_{ij}}
 \left[\frac{(E_i-E_j)-\nu_{ij}^{expt}}
            {\sigma_{\nu_{ij}^{expt}}}\right]^2. \label{neviazka}
\end{equation}

The values $\sigma_{\nu_{ij}^{expt}}$ are the RMS estimates of experimental 
errors (one standard deviation --- SD) for each experimental datum, and the 
summation is performed over all available experimental data. Due to the 
linearity of the equations (\ref{ridbergritz2}), the optimization problem 
comes to solving a system of linear algebraic equations. The complete 
solution involves the inversion of the $n_E * n_E$ sparse matrices. When 
the number of desired levels $n_E$ is of the order of a thousand, as is 
usually the case, this problem can be solved even using modern personal 
computers. (see \cite{LavrovRiazanovOiS} for details)). If the 
experimental errors are random and distributed according to a normal 
(Gaussian) law, the obtained solution corresponds to the maximum 
likelihood principle \cite{Hudson}.

The attempts to apply such statistical approach for determination of the 
energy level values appeared almost simultaneously in atomic (e. g. the 
spectra of the Er~I, Er~II \cite{VanderSluis}, Cl~II \cite{RK1974} by the 
inversion method, and the spectra of the Si~I \cite{RA1965}, 
Cl~I \cite{RK1969} by the iterative method) 
and in molecular \cite{BredohlHerzberg1973, Aslund1965}
spectroscopy when first 
digital computers became available. It is interesting to note, that in spite 
of obvious disadvantages of the iterative method (the problems of convergence 
and lack of the covariance matrix (see p.6 in \cite {LosAlamos1970}), its usage 
can be met in current publications (see e.g. \cite {Wetal2007}). 

In several studies reviewed by \AA slund in \cite{Aslund1965}, it was proposed 
to determine the rovibronic levels of molecules on the basis of the combination principle by 
solving the overdetermined system of equations with the use of a computer. 
Owing to the limited capabilities of the computers of the day, the optimization 
procedure was realized for processing the wavenumbers for separate bands only. 
As a result, those authors overlooked the opportunity to overcome the problem of 
the limitations for the method of common differences caused by the Laporte's 
selection rule allowing rovibronic transitions only between levels with 
different + and - parities \cite{LavrovRiazanovJetf}.  (Only recently, it was 
shown that, if it is possible to use experimental data on rovibronic lines 
that pairwise couple \it three or more \rm different electronic-vibrational 
states, then the system of equations (\ref{ridbergritz}) contains the link 
between the line wavenumbers and the values of all rovibronic levels involved 
\cite{LavrovRiazanovJetf}. Thus, in the case of heteronuclear molecules the 
above problem of the existence of uncoupled sets of the levels with even and 
odd rotational quantum numbers disappears.) Consequently, the relative positions 
of unrelated odd and even rotational levels, as in traditional approaches, were 
obtained using the Dunham approximation and molecular constants. It should be 
noted that the attempts of direct empirical determining the rovibronic levels 
in \cite{Aslund1965} were inconsistent from the very outset. The level values 
were considered as some intermediate parameters of a molecule, and the procedure
of their determination was only one of the steps on the long way from the 
measurement of the line positions on a photographic plate up to obtaining 
a particular set of vibrational-rotational constants for various electronic 
states. Later on, \AA slund \cite{Aslund1974} actually abandoned his initial 
idea in favour of so-called \it direct approach \rm \cite{Albriton} based on 
the one-step optimization of sets of vibro-rotational constants for upper and 
lower electronic states of a bad system.

The empirical determination of the energy level values by means of the least 
squares method looks natural and simple, but its correct realization in practice 
encounters serious difficulties. The matter is that the result of the minimization 
of (\ref{neviazka}) (i. e. the optimal set of the level values and the matrix of 
co-variances) depends not only on the values of experimental data, but also 
on weighting of various data (i. e. from what values of the error estimates 
$\sigma_{\nu_{ij}^{expt}}$ are included into the input data set). Therefore,
for the correct optimization of the adjusted parameters (required level values) 
it is necessary to know a dependable RMS error estimate for each experimental 
wavenumber 
value, or to have the set of the wavenumbers consisting from several subsets 
(statistical samples) of uniformly precise data with known RMS error estimates 
for each sample.

Unfortunately, authors of experimental works usually limit themselves to some 
remarks of general character (concerning resolving power of a spectrograph, 
linear dispertion, typical line widths, \it etc.\rm) and, at the best, to some 
rough estimates of an order of magnitude or the upper limit of possible experimental 
uncertainty. Most often the real accuracy of the reported data is uncertain. The 
situation in the literature containing experimental data analyzed in the present 
work will be illustrated below, but it should be mentioned that the same uncertainty 
may be met even in current publications. (To not be unsubstantiated, very recent 
paper \cite{Wetal2007} may be used as a typical example. In the part describing 
experimental setup one may read: "The estimated error is +0.005 \AA~for single lines, 
but many lines are blended in complex emission peaks."; while in the part describing 
data processing one may find: "As input to the code, 1314 classified lines were used 
with uncertainties on their wave numbers smoothly decreasing from 0.33 to 0.10 cm$^{-1}$ 
between 1150 and 2800 \AA." Attentive reader can easily recognize that the wavenumber 
uncertainties correspond to wavelength uncertainties 0.0044 and 0.0078 \AA~not equal 
to those for single lines. But a lot of questions remain. What about blended lines? 
Why the weighting of various data depends only on wavenumber values? What kind of the 
smooth decrease was chosen and why? We hope that the authors have certain answers and 
did their best to get best possible level values they are interested in. But what an 
independent researcher can be able to do with those wavenumbers after several decades?)

The problem of weighting various experimental data according to their accuracy in the 
framework of the least squares fitting was mentioned from the very beginning 
\cite{VanderSluis}. But in practical applications it was always considered as an issue 
of secondary importance, leaving a lot of room for author's subjectivity. May be it is 
because researchers use to process their own experimental data and have certain opinions 
concerning their accuracy. Then, irrespective of the way of the data processing, the 
empirical level values are considered as optimal when they are able to reproduce 
experimental wavenumbers with an accuracy corresponding to author's estimates for the 
experimental errors ("chosen tolerance"). That is certainly reasonable, but only if author's 
expectations coincide with real experimental uncertainty caused by all possible sources 
of errors. Thus, in spite of principle objectivity of the least squares method the 
results of its application have subjective character reflecting author's individuality. 
Achilles' heel of all known to authors of the present work papers devoted to the 
determination of the level values by means of the least squares adjustment is ignoring 
or subjective estimating of the experimental errors.

Development of pure statistical approach to the problem of empirical determination of 
energy level values \cite{LavrovRiazanovJetf} was motivated by author's desire to 
overcome the subjectivity discussed above. Main ideas of the approach may be formulated 
in the following way. Since optimal values of rovibronic levels are to be entirely 
defined only by amount and quality of available experimental data, a procedure of their 
obtaining should be based only on the fundamental principles. Rydberg-Ritz combination 
principle provides the link between experimental wavenumbers of rovibronic transitions 
and desired energy levels. Maximum likelihood principle gives the optimization criterion 
in the form (\ref{neviazka}) for the normal distribution function of random errors. 
The set of input data for minimizing (\ref{neviazka}) should be prepared by means of 
statistical analysis: 1) The data having systematic errors and spurious results 
(outliers) should be uncovered and eliminated; 2) The whole set of available experimental 
data should be divided into finite number of subsets (samples) of uniformly precise 
data with normal error distributions; 3) the values of an experimental RMS uncertainty for 
the samples (unique for all data belonging to the same sample) are obtained, being 
independent from those named in original papers. Optimal set of the rovibronic level 
values is obtained from the prepared input data set in the one-stage optimization 
procedure. The accuracy of empirical determination of the optimal level values is 
characterized by the covariance matrix $D(E)$, square roots of its diagonal elements are 
used as the uncertainties (one SD) of the level values ($\sigma_{E_i} = \sqrt{D_{ii}(E)}$). 
(The method may be easily 
generalized for other types of the distribution function of random errors.)

The computer code developed for practical realization of the method is based on the 
minimization of the weighted mean-square deviation (\ref{neviazka}) by the inversion 
of the $n_E * n_E$ sparse matrices \cite{LavrovRiazanovJetf,LavrovRiazanovOiS}. The $\nu_{ij}^{expt}$ 
and $\sigma_{\nu_{ij}^{expt}}$ values are the input data, while the optimal set of the 
energy levels $E_i$ and their covariance matrix $D(E)$ are the output data of the 
minimization. The later is used for calculating and visualization of various quantities 
suitable for interactive statistical analysis of the experimental data including 
rational censoring of various groups (samples) of the data and the derivation of the 
RMS uncertainties of experimental data independent from those reported in original papers.

The interactive analysis consists in the studies of the shape of the distribution 
functions of the weighted unbiased deviations 
\begin{equation}
\xi_{ij} =
  \frac{\nu_{ij}^{expt} -
        (\tilde E_i-\tilde E_j)}
       {\sigma_{\nu k}},\label{xi}
\end{equation}                             
for various sets (statistical samples) of the experimental wavenumbers, 
$k$ being the number of the sample.
Here the $\nu_{ij}^{expt}$ is certain experimental datum for the
wavenumber of the $i$ - $j$ transition, while the level values 
$\tilde E_i$ and $\tilde E_j$ are obtained by the minimization of 
(\ref{neviazka}) without using the particular datum under the
consideration. The $\sigma_{\nu k}$ is adjustable value of the 
estimate for experimental uncertainty (RMS error estimate) common for all the data 
included into the sample. The dimensionless deviation $\xi_{ij}$ characterizes 
a degree of co-ordination of the particular experimental datum $\nu_{ij}^{expt}$ 
with all other experimental data in the framework of the Rydberg-Ritz principle. 
Among all, the most important are certainly the data obtained for the same rovibronic 
transition in various works, and the wavenumbers of other lines directly connected 
with the upper n', v', N' or the lower n'', v'', N'' rovibronic levels.

The values $\tilde E_i$ and $\tilde E_j$ may be expressed through
the corresponding level values and the covariance matrix $D(E)$ obtained with the 
use of all experimental data, then

\begin{equation}
\xi_{ij} =
  \frac{\nu_{ij}^{expt} - (E_i - E_j)}
       {\sqrt{\sigma_{\nu k}^2 - (D_{ii}(E) + D_{jj}(E) - 2D_{ij}(E))}},
\end{equation}                             
where $D_{ij}(E)$ are elements of the covariance matrix. This makes it possible to 
obviate the need for the repeated minimizing (\ref{neviazka}).

The empirical accumulative distribution function for each data sample is defined by

\begin{equation}
F(\xi) = \frac{1}{n_{\nu}} \sum_{\xi_{ij}} I(\xi_{ij} \leq \xi),
\end{equation}                             
where $n_{\nu}$ is the number of elements in the sample,
$I(A)$ --- the indicator of event $A$.\footnote{$I(\xi_{ij} \leq \xi) = 1$ 
for $\xi_{ij} \leq \xi$ and $I(\xi_{ij} \leq \xi) = 0$ for $\xi_{ij} > \xi$.}
Here the sum is over all data included into the sample. 
After each minimization of (3) it is possible to calculate and to see
$F(\xi)$ for any sample of the $\xi_{ij}$ values chosen by the user of the 
computer code. That gives an opportunity to study various data samples
selected from the complete set of all experimental data
 by some physical considerations: the data reported in the certain paper,
wavenumbers of lines from the certain interval of wavelengths, or 
belonging to the certain band, or to the certain band system.

It is evident that if all the data $\nu_{ij}^{expt}$ included into some sample are free 
from systematic errors and have normal distribution of random errors with RMS estimate 
equal to $\sigma_{\nu k}$, then the empirical distribution $F(\xi)$ should be 
close to normal cumulative distribution function with the zero mean and 
variance equal to unity, namely

\begin{equation}
F_{0}(\xi) = \frac{1}{\sqrt{2 \pi}}
  \int\limits_{-\infty}^{\xi} exp(-\frac{x^2}{2}) dx.\label{norm_f}
\end{equation}

Therefore, it is useful and convenient to provide the interactive analysis of 
experimental wavenumber values by studies of the shape of the empirical functions

\begin{equation}
\Phi(\xi) = F_0^{-1}(F(\xi)), \label{Phi}
\end{equation}
calculated for various groups (samples) of available data.
Here $F_0^{-1}$ is the function reverse to $F_0(\xi)$, i.e.
$F_0^{-1}(F_0(\xi)) \equiv \xi$.

If the empirical distribution function $F(\xi)$ is close to normal distribution, 
the empirical function (\ref{Phi}) is close to linear function
 
\begin{equation}
\Phi(\xi) = \xi. \label{criterion}
\end{equation}                             

The way of dividing (sampling) the whole set of experimental data
onto the subsets is not determined leaving some subjectivity for a
researcher. But the shape of the function $\Phi(\xi)$ is objective 
characteristic of the sample being determined by experimental values 
of the wavenumbers, 
Rydberg-Ritz principal and only one arbitrary parameter $\sigma_{\nu k}$.
If for a certain data sample adjusting the value of the $\sigma_{\nu k}$
leads to the shape of $\Phi(\xi)$ fulfilling the criterion (\ref{criterion}),
the data included into the sample are to be considered as measured with uniform 
precision and characterized by the normal distribution of random experimental 
errors. The obtained value $\sigma_{\nu k}$ is their experimental
uncertainty (RMS error estimate). Thus, the method of \it a~posteriory \rm statistical
analysis proposed in \cite{LavrovRiazanovJetf, LavrovRiazanovOiS} 
gives the objective mean to get proper results by
the studies of the $\Phi(\xi)$ functions, trying various kinds of sampling and
adopting or rejecting various guesses.
First applications of the approach in our recent studies of rovibronic spectra of 
$BH$ and $H_2$ molecules 
\cite{LavrovRiazanovJetf, LavrovRiazanovOiS, BHterms, ALMU2008} showed that the method 
allows revealing and rejecting both single outliers (caused by wrong assignments, 
blending, spurious readings and misprints) and large data sets, including data of 
some experiments (probably caused by systematic error of the wavelengths 
calibration \cite{LavrovRiazanovOiS}).

\section{Statistical analysis of published wavenumbers of triplet 
rovibronic transitions}

All available values of rovibronic transition wavenumbers studied in emission, 
absorption, laser and anticrossing spectroscopic experiments of various authors 
\cite{Crosswhite, DabrHerz, Davies, DiekeBlue, Dieke, GloersenDieke, DiekePorto, 
FreundMiller} were analyzed in the present work. In overwhelming majority of the 
works the fine and hyperfine structures of lines have not been resolved, and the 
reported experimental wavenumbers correspond to intensity maxima of the 
observable line profiles. In cases of partly resolved structure of lines 
\cite{DabrHerz, Davies} we used wavenumbers of the brightest components.

The all identified triplet electronic transitions are naturally divided into three 
groups of the band systems. These are: 
i) the 9 band systems having one common low electronic state ($n^3\Lambda_g - a^3\Sigma_g^+$, 
with $\Lambda = 0, 1$ and $n = 3 - 9$); 
ii) the 12 band systems having another common low state ($n^3\Lambda_u - c^3\Pi_u$, with 
$\Lambda = 0 - 2$ and $n = 3 - 9$); and 
iii) the 7 band systems connecting  various $n^3\Lambda_g$ and $n^3\Lambda_u$ 
electronic states (with $\Lambda = 0 - 2$ and $n = 2 - 4$) including the 
$a^3\Sigma_g^+ - c^3\Pi_u$ transitions. 
Therefore, the whole set of available experimental data possesses different 
informational content about rovibronic levels of different electronic states. 
Thus, the information on rovibronic levels of the $c^3\Pi_u$ electronic state 
is contained in the wavenumber values of the 13 band systems, about the 
$a^3\Sigma_g^+$ state --- 10 band systems, and about for example the 
$e^3\Sigma_u^+$  state --- 6 band systems. For some electronic states 
as a source of the information about rovibronic levels we can use wavenumbers 
of lines from very few or even only one band system. In such cases the obtained 
values of the rovibronic levels are essentially less reliable. 

Most of the data used in our statistical analysis are those collected from earlier 
works (partly unpublished) and reported in \cite{Crosswhite}. 
This compilation contains wavenumbers of 
3117 rovibronic spectral lines assigned as triplet-to-triplet transitions. 31 from 
them have unassigned upper electronic and vibrational states. They were not used in 
the analysis, so only \it3086 \rm experimental data were taken from \cite{Crosswhite}. 
The 83 wavenumber values obtained in \cite{GloersenDieke} coincide with those reported 
in \cite{Crosswhite} (although \it12 \rm of them appeared in Appendix C of
\cite{Crosswhite} as unassigned). To prevent doubling of the same experimental results 
we excluded 71 wavenumbers from the data sets of \cite{GloersenDieke}. 
The data reported in \cite{DiekeBlue, Dieke} and 
\cite{Crosswhite} are obtained for the lines of the same band systems, but wavenumber 
values are different. Taking into account random character of the differences we 
decided to consider the data of those papers as the results of independent experiments. 
Thus, the initial data set used for the start of the statistical analysis contains: 
\it350 \rm wavenumbers from paper \cite{Dieke}, \it285 \rm from \cite{DiekeBlue}, 
all \it37 \rm  data from \cite{DiekePorto}, \it12 \rm  from \cite{GloersenDieke}, 
\it1 \rm from \cite{FreundMiller}, \it3086 \rm from \cite{Crosswhite}, \it81 \rm 
from \cite{DabrHerz}, and \it3 \rm from \cite{Davies}. It should be stressed that 
the wavenumbers of spectral lines used in the present work are spread over the very
wide range 0.896---28166.84 cm$^{-1}$ from radio frequencies up to the ultraviolet.

\begin{figure}[ht!]
\begin{center}
\epsfig{file=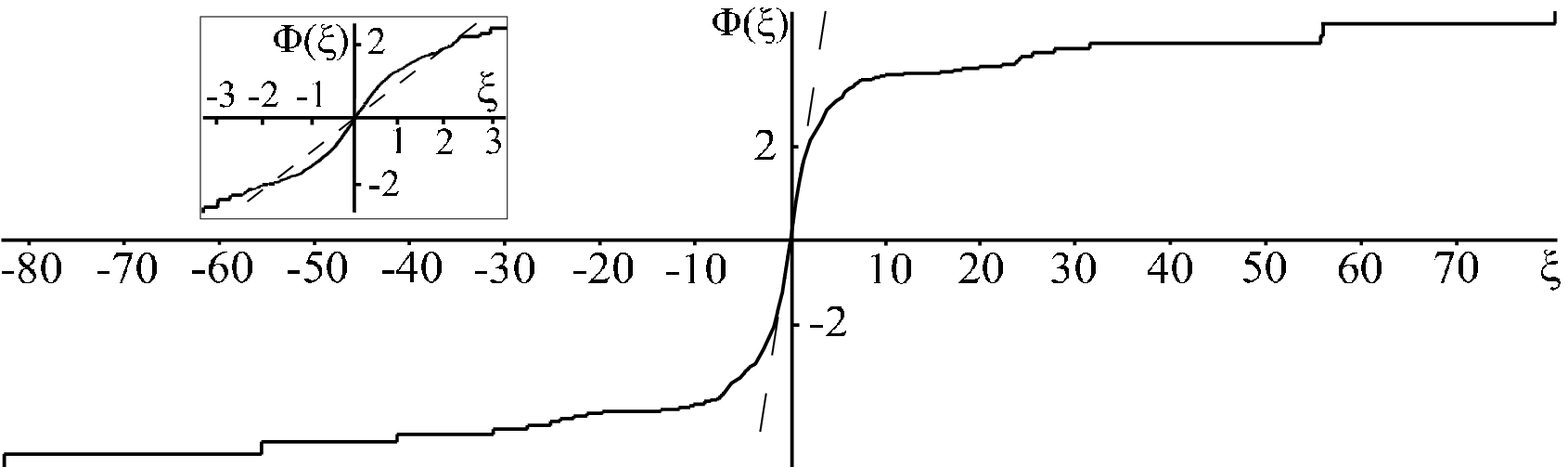, width=0.8\columnwidth,clip}
\end{center}
\caption{Empirical function $\Phi(\xi)$ obtained when all published experimental data with 
RMS uncertainties equal to the error estimates reported in original papers. The onset 
represents the central part of the $\Phi(\xi)$ in larger scale.}\label{initdistr}
\end{figure}

On the first iteration of our statistical analysis we assumed, that all
sets of experimental data presented in various papers 
represent the samples of uniformly precise data with RMS error uncertainty
equal to the estimates presented in original experimental works. 
In the review \cite{Crosswhite} there are two remarks concerning accuracy of
the wavenumbers. In experimental part it is written: "... the widths of the lines 
themselves limited the attainable accuracy, which was a few hundreds of a cm$^{-1}$".
At the end of the analysis the authors of \cite{Crosswhite} mentioned: "The mean (O-C)
value for all triplets is $-0.001 \pm 0.051$ cm$^{-1}$", and came to the conclusion that
"The precision of the measurements appear to be roughly uniform from the infrared
to the ultraviolet". In spite of doubtfulness of such estimations we have accepted
the value $\sigma_{\nu} = 0.05$ cm$^{-1}$ in the first iteration of our analysis for
all wavenumbers taken from \cite{Crosswhite}.
In papers \cite{DiekeBlue, Dieke, DiekePorto} experimental
errors are not mentioned at all. Therefore, we used the same value 0.05 cm$^{-1}$ 
taking into account the estimates made by the same scientific group.
In \cite{GloersenDieke} it is written: "... the errors in the wave number of good
lines should not exceed a few times 0.01 cm$^{-1}$". We certainly don't know which
lines are good or bad, therefore in this case we again took $\sigma_{\nu} = 0.05$ cm$^{-1}$.
For the data taken from \cite{DabrHerz, Davies, FreundMiller} 
we used author's estimates $\sigma_{\nu} = 0.01, 0.003, 0.0003$ cm$^{-1}$, respectively.
Under these assumptions we obtained a set of
energy levels by minimization of (\ref{neviazka}) 
and the empirical function $\Phi(\xi)$ shown in figure~\ref{initdistr}. 

One may see that the distribution function $F(\xi)$ is far from the normal
distribution $F_0(\xi)$ ($\Phi(\xi)$ is too far from the dotted straight line
representing the case of normal distribution). 
Too many experimental wavenumber values significantly exceed those calculated as
differences of corresponding energy level values. The distribution is not close to 
normal even in the area of small deviations ($\xi < 3$). We suppose that this 
disagreement is caused by the deviations of real experimental wavenumber 
errors and their estimates, reported in original works. 
(On this stage of our analysis it was observed that 33 lines with wavenumbers 
measured in \cite{Crosswhite}
are the combinations of 55 rovibronic levels representing the blocks of levels
totally disconnected with all other triplet rovibronic levels of $D_2$. 
Therefore this wavenumbers were excluded from further consideration as 
useless. Therefore, only \it3053 \rm from the data reported in \cite{Crosswhite}
are useful for further analysis.)

Thus there is the necessity of independent estimating of experimental 
uncertainties of the measured wavenumbers. To carry out this estimation 
we need to divide the data set used in the first iteration into some groups 
(subsets, samples) so, that
all experimental data inside one subset can be considered as uniformly precise. 
We tried to sort experimental data to subsets 
taking into account various criteria, selecting the data obtained in various papers, in
various wavelength regions and in various band systems. But acceptable results for error 
estimates were obtained only when we considered each band as a group of uniformly precise data. 
May be that way of sampling is fruitful because the lines of each band are located in narrow
parts of the spectrum, and different bands are usually measured and treated by
different people even within the same scientific group. It should be also mentioned, that
such approach gave reasonable results in similar analysis of $BH$ \cite{BHterms} and
$H_2$ \cite{ALMU2008} spectra.

Each m-th iteration consists of: 
1) the minimization of (\ref{neviazka}) with certain m-th set of the input data prepared at the
last stage of the previous (m-1)-th iteration. It includes certain amount of experimental wavenumbers
divided into finite number of samples characterized by certain RMS estimates for experimental
errors in each k-th sample $\sigma_{\nu k}^{(m - 1)}$;
2) the output data (optimal energy levels $E_i^{(m)}$ and the covariance matrix $D_{ij}^{(m)}$)
are used for calculating the $\Phi^{(m)}(\xi)$ functions for each sample. That gives us an 
opportunity to analyze the shapes of the $F_k^{(m)}(\xi)$ distribution function for all the 
samples and to find outliers. The data with $\xi_{ij}^{(m)}$ larger than 3 were
qualified as outliers. In all cases this criterion of censoring
was less strong then commonly adopted Chauvenet's and Peirce's criterions \cite{Ross2003}; 
3) The new set of input data for the next (m+1)-th iteration is prepared by excluding 
the outliers from current (m)-th data set, and by obtaining for each sample new values
of RMS estimates for experimental errors $\sigma_{\nu k}^{(m)}$ by multiplying the 
$\sigma_{\nu k}^{(m - 1)}$
by the factor which should move the $\sigma_{\xi k}$ closer to unity on the next iteration. 
Usually we used the expression 
$\sigma_{\nu k}^{(m)} = \sigma_{\nu k}^{(m - 1)} \sigma_{\xi k}^{(m)}$,
which is not strict due to the nonlinear link between $\sigma_{\nu k}^{(m)}$ and
$\sigma_{\xi k}^{(m + 1)}$.

The square roots of the second moments $\sigma_{\xi k}^{(m)}$ for 
empirical distribution functions $F_k^{(m)}(\xi)$ (for each sample in each iteration) 
are calculated as

\begin{equation}
\sigma_{\xi k}^{(m)} = \sqrt{\sum_{\xi_{ij}^{(m)}} (\xi_{ij}^{(m)} - <\xi_{ij}^{(m)}>)^2 F_k^{(m)}(\xi_{ij}^{(m)})},
\end{equation}
where $<\xi_{ij}^{(m)}> = \sum \limits_{\xi_{ij}^{(m)}} \xi_{ij}^{(m)} F_k^{(m)}(\xi_{ij}^{(m)})$, and 
the summing is performed over all the data included into the $k$-th sample.

The method of independent error estimation of wavenumbers belonging to the same $k$-th sample
may be illustrated by considering one typical
example --- analysis of the $(0-1)$ band of the $e^3\Sigma_u^+ - a^3\Sigma_g^+$
electronic transition.
30 spectral line wavenumber values for $R$- and $P$-branch lines from this band are reported 
in \cite{Crosswhite}. 
The lines $R18$ and $R19$ are caused by the transitions between the 
($E_{e, 0, 19}$, $E_{a, 1, 18}$) and ($E_{e, 0, 20}$, $E_{a, 1, 19}$) pairs of
rovibronic levels, respectively.
Mutual positions of these levels
can be derived using wavenumber values of these lines only, because 
there are no other lines, which
provide information about these levels. Then the value (\ref{xi}) for these lines is equal 
to zero. We don't take into account such
lines when calculating empirical distribution functions $\Phi_k^{(m)}(\xi)$. 

\begin{figure}[ht!]
\begin{center}
\epsfig{file=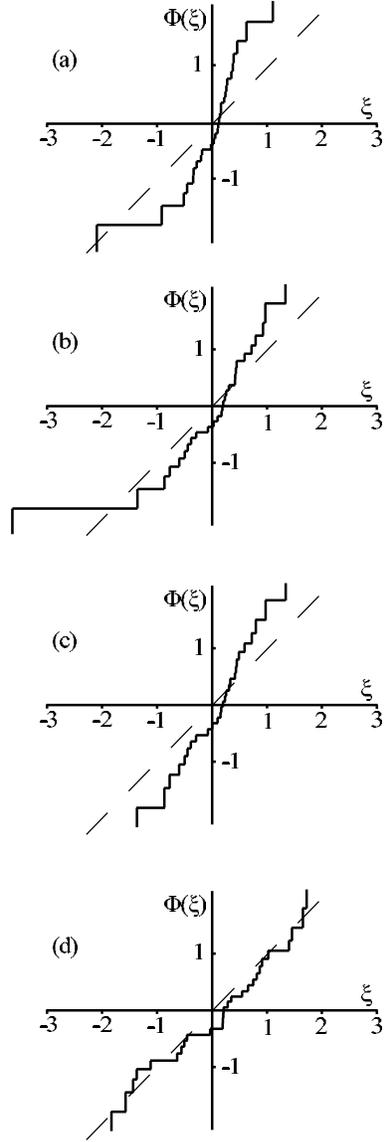, width=0.3\columnwidth,clip}
\end{center}
\caption{Empirical functions $\Phi^{(m)}(\xi)$ obtained in different m-th 
iterations (a, b, c, and d correspond to $m$ = 1, 2, 3, and 4)
of the statistical analysis of the sample of experimental data 
representing wavenumbers of R- and P- branch lines of the
$(0-1)$ band of the $e^3\Sigma_u^+ - a^3\Sigma_g^+$ electronic 
transition of $D_2$.}\label{bandworkdistr}
\end{figure}

The empirical function $\Phi_k^{(1)}(\xi)$ obtained during first iteration 
(based on the sample of remaining 28 experimental wavenumbers 
characterized by common estimate $\sigma_{\nu k}^{(0)} = 0.05$ cm$^{-1}$ 
mentioned in \cite{Crosswhite}) is presented in figure~\ref{bandworkdistr}(a).
One may see, that the distribution $F_k^{(1)}(\xi)$ is far from the normal
distribution, 
which is represented by a dotted line on the figure. 
The square root of the second moment of this distribution 
$\sigma_{\xi k}^{(1)} = 0.56 \pm 0.14 = 0.56(14)$ is
not close to unity. We may try to shift the $\sigma_{\xi k}^{(2)}$ to unity by
changing error estimation in the following way:
$\sigma_{\nu k}^{(1)} = \sigma_{\nu k}^{(0)} \sigma_{\xi k}^{(1)} = 0.028(8)$ cm$^{-1}$.

In the second iteration for the band under the consideration we used the same set of 
experimental data but new value of $\sigma_{\nu k}^{(1)}$ instead of $\sigma_{\nu k}^{(0)}$. 
The empirical function
$\Phi_k^{(2)}$ is shown in figure~\ref{bandworkdistr}(b). One may see that now the majority 
of lines shows the deviations sufficiently closer to the normal distribution, and
the square root of the second moment $\sigma_{\xi k}^{(2)} = 0.9(3)$ is
closer to unity. But one point located in the bottom
left corner of the figure is in contradiction with rest of the distribution.
This point represents the $\xi$ value for
the line $P10$. The measured wavenumber value of this line 
$\nu_{a, 1, 10}^{e, 0, 9} = 9292.16$ cm$^{-1}$ deviates from the difference of 
corresponding energy level values derived in this iteration
$\nu_{a, 1, 10}^{e, 0, 9} - (E_{e, 0, 9}^{(2)} - E_{a, 1, 10}^{(2)}) = -0.0950$~cm$^{-1}$. 
The deviation exceeds our estimate for RMS error
$\sigma_{\nu k}^{(1)} = 0.028(8)$ cm$^{-1}$ more than three times.
Thus the experimental wavenumber for the $P10$ line should be excluded from
the input data set as the outlier.

In the third iteration we used 27 wavenumber values and common uncertainty
$\sigma_{\nu k}^{(2)} = \sigma_{\nu k}^{(1)} = 0.028(8)$ cm$^{-1}$. The
function $\Phi_k^{(3)}(\xi)$ thus obtained is shown in figure~\ref{bandworkdistr}(c).
One may see that this distribution function
is not close to normal distribution function mainly because $\sigma_{\xi 
k}^{(3)} = 0.60(8)$ is too far from unity. Therefore we have to correct 
the experimental uncertainty as
$\sigma_{\nu k}^{(3)} = \sigma_{\nu k}^{(2)} \sigma_{\xi k}^{(3)}$. After certain number
of iterations we are coming to $\sigma_{\nu k} = 0.012(1)$ cm$^{-1}$. Then on
the next iteration we have empirical $\Phi_k(\xi)$ function shown in figure~\ref{bandworkdistr}(d)
and the square root of the second moment $\sigma_{\xi k} = 1.0(1)$.
One may see that now the experimental distribution function $F_k(\xi)$ 
is close to the normal distribution function with zero mean and the
 variance equal to unity.

Thus, the sample of the 27 wavenumbers under the study may be considered 
as a group of uniformly precise experimental data with RMS estimate for
random errors $\sigma_{\nu k} = 0.012(1)$ cm$^{-1}$. It should be underlined
that our estimate is less than a quarter of the value $0.05$ cm$^{-1}$
declared in \cite{Crosswhite}. 

In that way we estimated RMS uncertainties of experimental data by the 
rational censoring of experimental data and determination of proper 
values of $\sigma_{\nu k}$. 

After several dozens of iterations described above we 
came to the conclusion that 234 experimental data representing 228 spectral lines have
to be exclude as outliers. Remaining 3588 wavenumbers were organized as 317 samples of
uniformly precise data with RMS uncertainties obtained by the statistical analysis. 
RMS errors for many bands are so close that it seams reasonable to 
provide the enlargement of samples by merging studied samples having close values of 
RMS error.
This also helps to improve statistics because some bands have small amount of data.
It is reasonable to unite into the same sample wavenumbers of various bands with
RMS errors getting to the certain wavenumber interval. 
The amount and widths of such intervals are uncertain. 
The choice of these parameters may lead to some subjectivity.
On one hand, the widths of intervals should be large enough for the amount of data
included in each interval be sufficient for statistics.
But on the other hand, it should be as small as possible, 
not to disturb the close to normal error distributions obtained for separate bands.
Our computational experiments show, that the intervals $(0.00\div0.01)$ cm$^{-1}$,
$(0.01\div0.02)$ cm$^{-1}$, $(0.02\div0.03)$ cm$^{-1}$, etc. fullfil 
the requirements mentioned above.
It is important, that all energy level values, 
obtained by the minimization of (\ref{neviazka}) with experimental data 
divided into the such enlarged samples of this size, have been changed 
significantly less then one SD of their 
determination with the original 317 samples.

After that, the whole set of the 3588 experimental data was reorganized 
into the 17 enlarged samples of uniformly precise data from 
\cite{Crosswhite, DabrHerz, DiekeBlue, Dieke, GloersenDieke, DiekePorto}.
The RMS error estimates for each of these 17 subsets were derived
by the method similar to described above. The results of the distribution 
of the data among the 17 samples together with the list of RMS error estimates 
for the wavenumbers of lines belonging to various vibronic bands is 
presented in table~1 of \cite{LU2007}. Also 1 wavenumber from 
\cite{FreundMiller} and 3 wavenumbers from \cite{Davies} were used in 
further analysis as two separate samples (18-th and 19-th) 
with RMS errors reported in original papers, because the errors were 
unchanged during the analysis. 

\begin{figure}[ht!]
\begin{center}
\epsfig{file=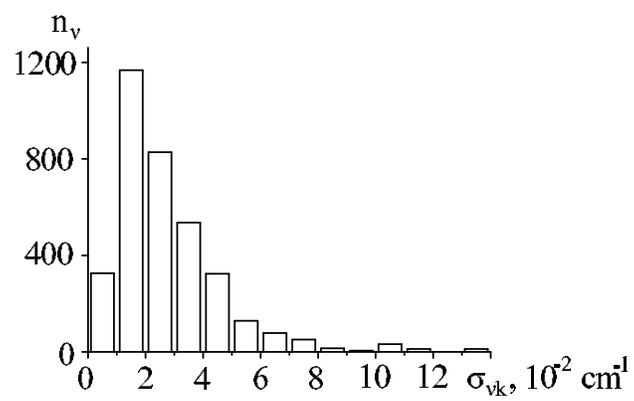, width=0.5\columnwidth,clip}
\end{center}
\caption{The amount of the wavenumber values $n_{\nu}$ within various enlarged samples of 
the uniformly precise experimental data.}\label{nnu}
\end{figure}

The amount of experimental data $n_{\nu}$ in each of the new samples is 
presented in figure~\ref{nnu}. It is seen, that the amount of data in each sample 
obtained by the enlargement is sufficient for carrying out the statistical 
analysis. Moreover, the distribution of data over the error steps looks quite 
plausible: the number of too precise data ($\sigma_{\nu k} = (0.00 \div 0.01$) cm$^{-1}$) 
is small ($\approx 9 \%$), most of the data ($\approx 85 \%$) are measured with realistic 
accuracy ($0.01 \div 0.06$) cm$^{-1}$, and the quantity of data monotonously falls 
down with further increase in the error estimate $\sigma_{\nu k}$.

\begin{figure}[ht!]
\begin{center}
\epsfig{file=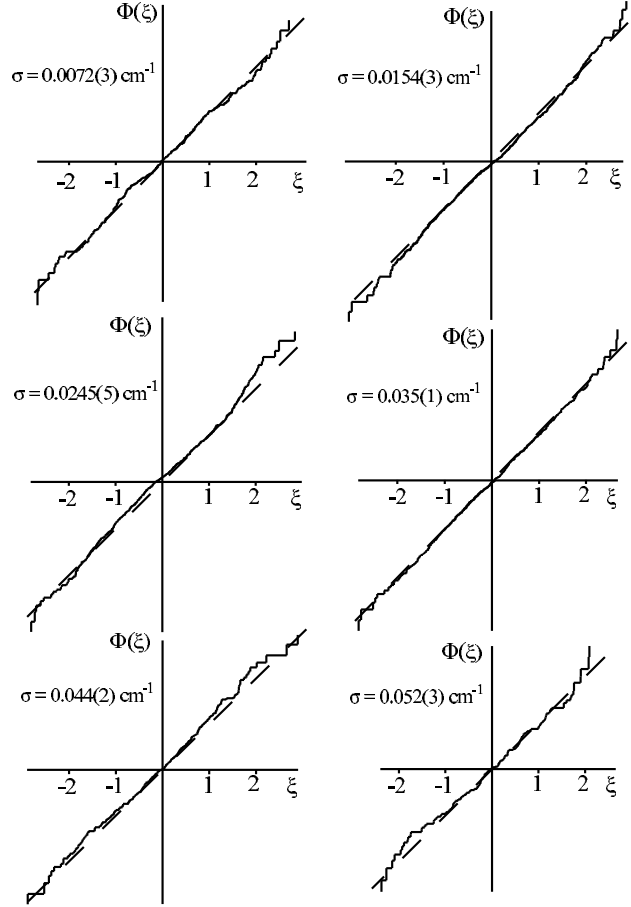, width=0.5\columnwidth,clip}
\end{center}
\caption{Empirical functions $\Phi(\xi)$ representing the error 
distributions within the first six enlarged samples corresponding 
to the intervals $(0 \div 0.01)$, $(0.01 \div 0.02)$, $(0.02 \div 0.03)$, 
$(0.03 \div 0.04)$, $(0.04 \div 0.05)$, $(0.05 \div 0.06)$ 
(from left to right, from top to bottom). 
The values of the experimental uncertainty obtained in the present work 
for every sample are shown near appropriate graphs together with SD of 
their statistical determination (in brackets).}\label{uniteddistr}
\end{figure}

Empirical functions $\Phi(\xi)$ for first six united samples representing
vast majority of useable experimental data are shown in figure~\ref{uniteddistr}.
Correspondent values of RMS error estimates for each sample are shown in the 
figure. One may see that the error distribution functions are in much better 
accordance with normal distribution function then those obtained for the data 
samples representing separate bands (see figure~\ref{bandworkdistr}).

Thus, splitting of experimental data into the samples and the subsequent 
enlargement of the samples as a whole were justified. Taking into account
high enough statistics of the samples we are coming to the conclusion that
obtained error estimates with a high probability are close to the values of
the real experimental uncertainty.

\section{Experimental determination of wavenumbers for questionable lines}

The statistical analysis of the 3822 known to authors published wavenumber
values has shown that 3588 of them are useable for empirical determination 
of the level values, while the 234 experimental data (concerning 228 spectral 
lines) have to be excluded as outliers.
161 of those questionable lines are
located within the wavelength range $4300 \div 7300$ \AA~ available for us. 
26 from them are the data reported in old \cite{DiekeBlue, Dieke} papers and
the wavenumber values for these lines were measured once more and corrected in
the compilation \cite{Crosswhite}.
For remaining 135 lines we decided
to provide independent experimental determination of their wavenumbers.
For that purpose we use the emission spectra of $D_2$ obtained during our
studies of translational and rotational temperatures in hydrogen and
deuterium containing plasmas \cite{LKOR1997}. Detailed description of the
experimental setup was reported elsewhere \cite{AKKKLOR1996}. Capillary arc
discharge lamps DDS-30 described in \cite{LT1982} have been used as a light
source. They were filled with about 6 Torr of the spectroscopically pure
$D_2$ + $H_2$ (9:l) mixture. The range of the discharge current was from
50 to 300 mA (current densities $j=1.6 \div 10$ $A/cm^2$). 
The light from the axis of plasma inside the capillary
was directly focused by an achromatic lens on the entrance slit of the
Czerny-Turner type l m double monochromator (Jobin Yvon, U1000).
The intensity distribution in the focal plane of the spectrometer was
recorded by cooled CCD matrix detector of the Optical Multichannel
Analyser (Princeton Appl. Res., OMA-Vision-CCD System).

\begin{figure}[ht!]
\begin{center}
\epsfig{file=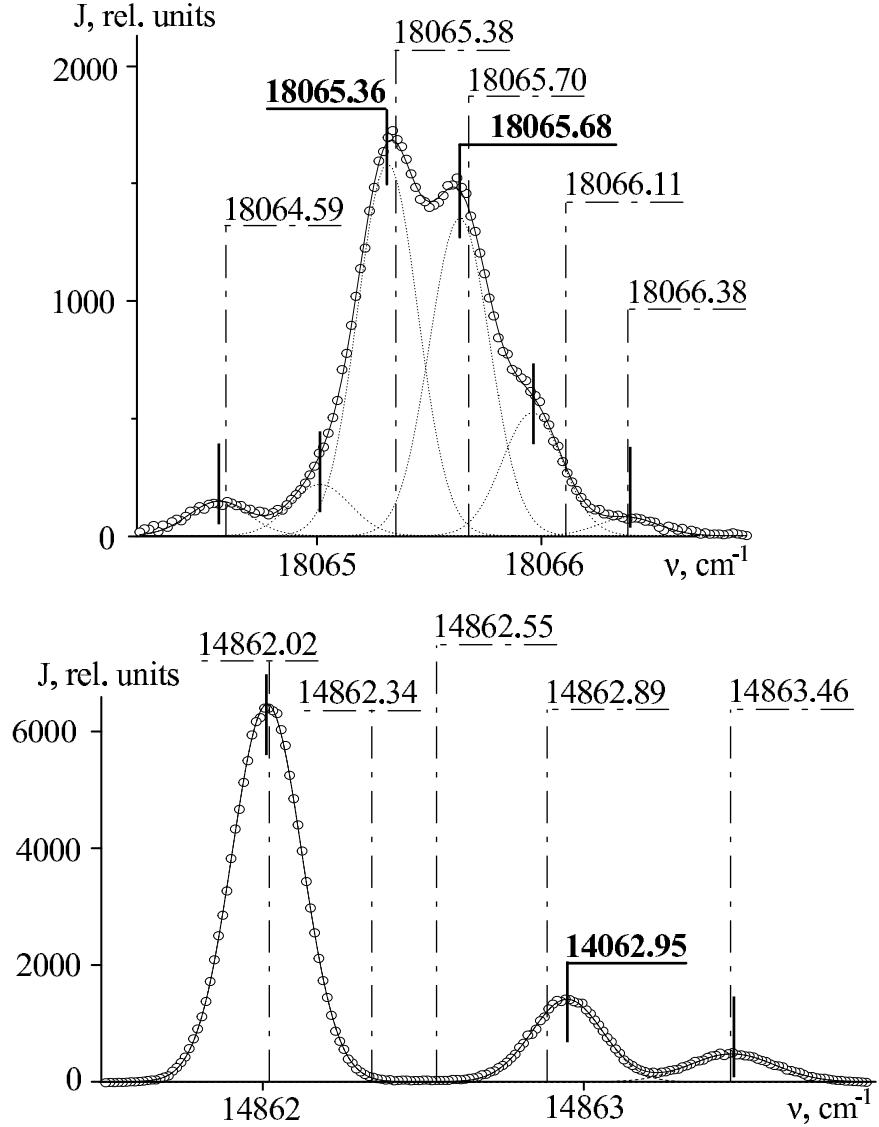, width=0.7\columnwidth,clip}
\end{center}
\caption{Fragments of the $D_2$ spectrum in the neighbourhood of
the questionable lines. Experimental intensity J in relative units is
shown by open circles. Dotted lines represent Gaussian 
profiles for separate lines obtained by deconvolution, while solid
line corresponds to the total intensity obtained by summing over
the components. The wavenumber values from the Appendix C of 
\cite{Crosswhite} are shown and marked by dash-dot vertical lines. 
Pieces of bold vertical lines denote maxima of the separate line profiles.
Bold underlined wavenumbers represent the new wavenumber values 
for the questionable lines obtained in the present work.}\label{spectrum}
\end{figure}

For identification of the $D_2$ spectral lines in our spectrum we used the
assignments and wavelength values from \cite{Crosswhite}. Those values show
certain spread around monotonic dependence of the wavenumber from the
distance along the direction of dispersion in the focal plane of the
spectrograph. The dispersion function of the spectrograph was obtained by the
polynomial least squares fitting of the wavelengths versus the distance
for strong unblended lines. Such wavelength calibration allow us to get new 
experimental values for the wavenumbers of 125 questionable spectral lines.
The new values differs from those reported in \cite{Crosswhite} not only
because of the smoothing procedure. In contrast to previous works we used
digital intensity registration providing linear response of the CCD detector.
That gave us the opportunity of digital deconvolution of the recorded spectra to
resolve the major part of blended lines. The observed line profiles were
determined mainly by Doppler and instrumental broadening.
For strong unblended lines they were close to Gaussian shape except of 
insignificant far wings 
(see e. g. bottom graph in figure~\ref{spectrum}). Therefore, the parts of spectra
in the neighbourhood of the blended questionable lines under the study were 
approximated by the superpositions of certain number of lines having Gaussian
profiles with fixed half width and adjustable intensity and wavelength values.
Then we obtained new values for the maximum intensities of $D_2$ lines.
Typical example is shown in the figure~\ref{spectrum}. One may see that the new
values of the rovibronic transition wavenumbers are a little bit different
from those reported in \cite{Crosswhite}.
In our experimental conditions the intensities of 10 questionable 
lines were too weak for detecting and unambiguous separation by the
deconvolution. 

\begin{figure}[ht!]
\begin{center}
\epsfig{file=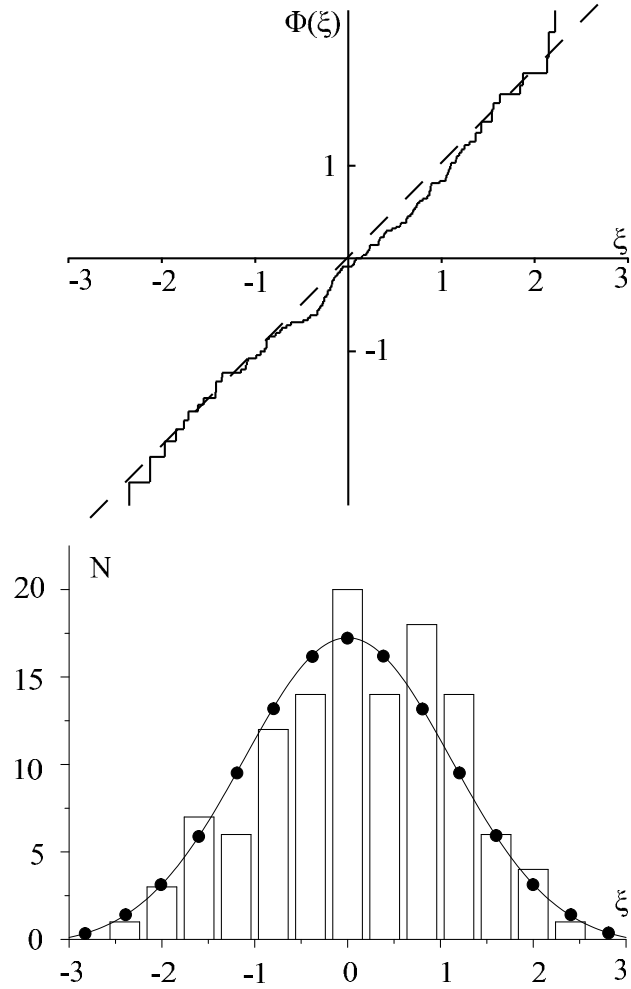, width=0.5\columnwidth,clip}
\end{center}
\caption{Empirical function $\Phi(\xi)$ for the sample containing
new wavenumber values
for questionable lines. The frequency diagram $N$ versus $\xi$ is constructed
with uniform width intervals equal to $0.4$. $N$ represents the number of 
the deviations $\xi_{ij}$ within each interval. The columns show experimental
results while the solid line with circles corresponds to the normal distribution
with zero mean and variance equal to unity.}\label{deletedlinesdistr}
\end{figure}

The uncertainty of our experimental data is mainly determined by
errors of the deconvolution process. It should be underlined, that 
most of the questionable lines are blended lines of relatively low intensity.
In that case it is not easy to obtain reliable RMS estimates for 
experimental errors. Therefore we used statistical method for 
derivation of experimental uncertainty described above. 
The new wavenumber values 
were included into the input data set obtained in the last stage of
previous section as a separate (20-th) sample of uniformly precise 
experimental data.
After several iterations the root mean square estimate of
experimental uncertainty $0.06$ cm$^{-1}$ was obtained without any evidence
for existence of outliers. The experimental $\Phi(\xi)$ function for 
the sample of new wavenumbers for questionable lines is shown in 
Figure \ref{deletedlinesdistr} together with corresponding frequency diagram. 
One may see that the distribution function of random errors is close to 
the normal distribution. Moreover new wavenumber values do not contradict
to the rest of the data set and may be used for determination of
optimal values of rovibronic levels.

The new wavenumber values of questionable lines obtained in the
present work are listed in table~\ref{NewLines} together with those
from \cite{Crosswhite}. One may see that in most cases the differences
between new and old data are significant. Most often small differences 
($ < 0.05$ cm$^{-1}$) are
probably caused by errors of reading from photographic plates and/or
by round-up errors, when the vacuum wavenumbers were calculated from the 
wavelengths measured in air. In our case they were eliminated by digital 
reading of the intensity profiles and smoothing procedure of the
determination of the dispersion curve.
The differences higher than $0.05$ cm$^{-1}$ are caused by blending (by shifts of 
the intensity maxima or by disappearance of a weak line within the profile of 
the strong one, see e.g. figure \ref{spectrum}), misprints and wrong assignments.
The differences (O-C) between Observed wavenumber values and those 
Calculated
as differences of corresponding optimal energy level values obtained in the 
present work are also shown in table~\ref{NewLines}. One may see that in the 
framework of 
Rydberg-Ritz principle the new experimental data for 125 questionable lines
are in much better co-ordination with the 3588 wavenumbers of other lines 
than those reported in \cite{Crosswhite}.

\section{Results and discussion}

Information on the amount of experimental data reported in
original papers and the amount of outliers revealed by statistical 
analysis is presented in table~\ref{LineNumber}. 

\begin{figure}[ht!]
\begin{center}
\epsfig{file=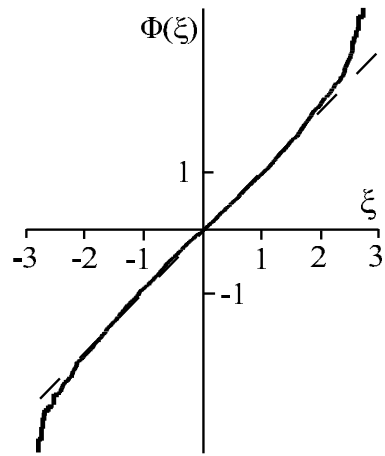, width=0.3\columnwidth,clip}
\end{center}
\caption{Empirical function $\Phi(\xi)$ obtained for the final set of all 
available experimental data (see the text) with RMS errors estimates obtained 
in the present work by the statistical analysis.}\label{finaldistr}
\end{figure}

After all, the input data set consists of 3713 wavenumber values
(3588 old and 125 new)
divided into 20 subsets of uniformly precise
data with known RMS estimates of experimental uncertainties
obtained in the present work by pure statistical analysis, i.e.
independent from the estimates reported in original papers. This
data set was used for determination of optimal values of all 
studied experimentally  triplet
rovibronic levels by minimizing (\ref{neviazka}). 
The empirical function $\Phi(\xi)$ thus obtained is presented in
figure \ref{finaldistr}. One may see that the final distribution 
function $F(\xi)$ is close to the normal 
distribution $F_0(\xi)$. Small deviations from the linear plot 
are caused by insufficient statistics for the data 
showing too high values of the weighted deviations $|\xi_{ij}|$
(without any contradiction with the Chauvenet's criterion).
Thus, optimal values of the 1050 rovibronic energy levels 
have been obtained 
from the 3713 values of the experimental wavenumbers. 
All the energy levels were obtained relative to the
lowest vibro-rotational level ($v = 0$, $N = 0$) of 
the $a^3\Sigma_g^+$ electronic state.

The shift ($33.631 \pm 0.004$) cm$^{-1}$ between uncoupled levels of ortho- 
and paradeuterium was obtained by the least squares analysis of the 
$a^3\Sigma_g^+$, $v = 0$, $N = 0 \div 18$ levels with odd and even 
values of the rotational quantum number $N$. According to 
\cite{Crosswhite} the difference 
$E_{a01} - E_{a00}$ = ($33.62 \pm 0.10$) cm$^{-1}$.
Both values coincide within error bars, but our value is obtained 
with much higher precision. Taking into account the values of the 
SD uncertainties of our optimal level values (see table~\ref{NewLevels}), 
this increase of precision in obtaining 
the shift value is significant. 

Optimal values of the rovibronic levels for the triplet electronic states
corresponding to principle quantum numbers $n = 2$, $3$ of the united atom
are listed in table~\ref{NewLevels}. Radiative transitions between those levels
are located in visible part of spectra, which is most convenient for 
spectroscopic studies of non-equilibrium gases and plasmas 
\cite{LMKR1999, RDKL2001}. The expanded version of the table~\ref{NewLevels} 
including all 1050 optimal level values is presented in \cite{LU2007}.
For each rovibronic level we introduced into the table~\ref{NewLevels} 
the standard deviation of the empirical determination (in brackets), 
the number of spectral lines $n_{\nu}$ which originate or terminate on 
that level, 
and the difference $\Delta E$ between energy level values reported in 
\cite{Crosswhite} and those obtained in the present work.

Several interesting things could be seen from the table~\ref{NewLevels}, namely:

i) In many cases the number of the lines, directly connected with 
certain level, is high enough for statistics. In some cases 
there are very few lines.
In that cases the levels are obtained with less accuracy. It is important to
stress, that the SD uncertainties of empirical determination of optimal level 
values partly reflect the influence of $n_{\nu}$ on the SD values. This is 
the trend (not straightforward dependence) because there are more 
factors of the influence. The levels with $n_{\nu} < 4$ should be considered 
as unreliable. The SD of such 
levels are mainly determined by experimental uncertainties of few lines 
directly connected with those levels. Additional experimental studies
are needed to obtain more reliable results.

ii) The differences $\Delta E$ are generally less, than the value
$0.05$ cm$^{-1}$, reported in \cite{Crosswhite} as SD uncertainty of the energy 
levels. This confirms our assumption that the
unique SD uncertainty for all the levels declared in \cite{Crosswhite} 
is only an upper-limit estimate of real uncertainties of various levels. 

iii) The vast majority of the $\Delta E$ are much higher 
than SD error bars of our optimal values, which normally
are within the range of $0.004 \div 0.03$ cm$^{-1}$ depending on
the value of rotational and vibrational quantum numbers. Thus, the 
deviations of the data reported in \cite{Crosswhite} 
from the optimal level values obtained in the present work are significant.

Currently it is not possible to find out and list reasons for those 
deviations for all and every levels, because \cite{Crosswhite} 
does not contain information about details of calculating 
the common differences in various
bands. We may presume that they appeared as a result of following reasons:
errors of reading from photo plates, round up errors in calculating
the wavenumbers from measured wavelengths, lack of deconvolution
of blended lines, subjective preferences in eliminating some
of controversial experimental data, some additional assumptions
(like neglecting $\Lambda$ --- doubling for some states), transfer and 
accumulation of experimental errors in the method of common differences
adopted in \cite{Crosswhite} for wavenumber analysis of separate bands, 
quite subjective multi-step procedure of the analysis of the band systems. 
It is clear that for various rovibronic states all these factors may 
multiply or compensate each other. 

\begin{figure}[ht!]
\begin{center}
\epsfig{file=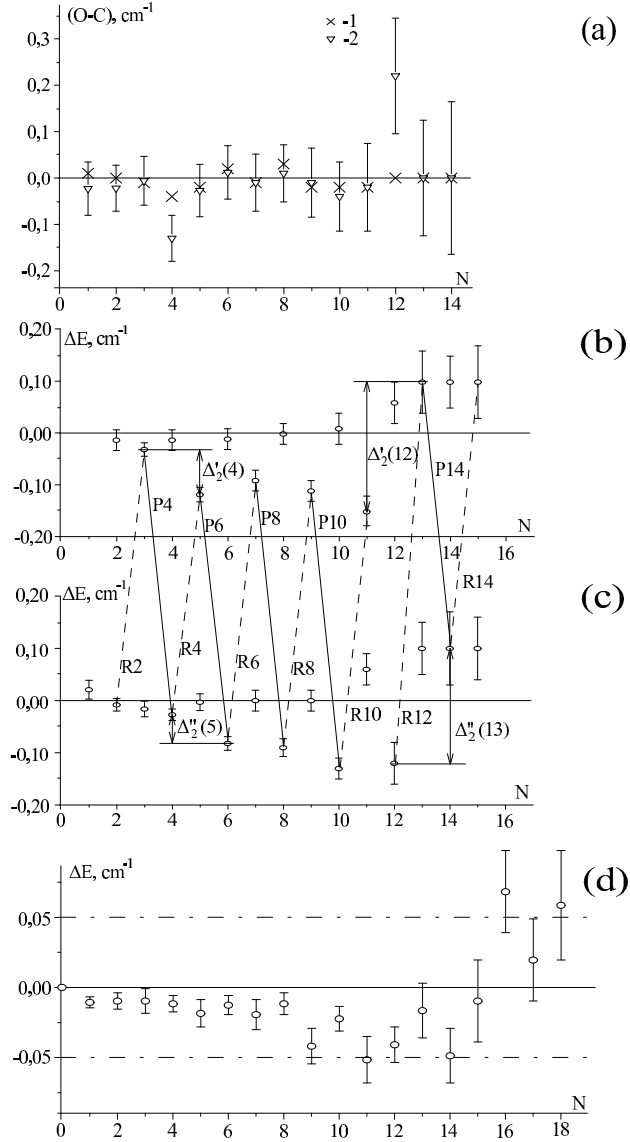, width=0.5\columnwidth,clip}
\end{center}
\caption{The differences $(O-C)$ between observed wavenumbers from 
\cite{Crosswhite} and those calculated as the differences of 
empirical levels from \cite{Crosswhite} (1) 
and of the optimal level values (2) for 
$R$-branch lines of the $(0-0)$ band of the 
$j^3\Delta_u^-$ --- $c^3\Pi_u^-$ electronic transition (a).
The error bars in figure (a) represent the sum of standard 
deviations for optimal values of upper and lower levels and 
for the wavenumber value (obtained in the present work for 
the set of experimental data from \cite{Crosswhite} belonging 
to this band).
The differences $\Delta E$ between the level values from
\cite{Crosswhite} and the optimal level values (Table~\ref{NewLevels})
for various rotational levels of the $j^3\Delta_u^-$, $v = 0$ (b); 
$c^3\Pi_u^-$, $v = 0$ (c); and $a^3\Sigma_g^+$, $v = 0$ (d) 
electronic-vibrational states of $D_2$ molecule.
Spectral lines of the $R$- and $P$- branches of the (0--0) band of the
$j^3\Delta_u^-$ --- $c^3\Pi_u^-$ electronic transition connecting some of the 
levels are shown by vertical lines (dashed for $R$-branch and solid for 
$P$-branch) with standard labels. The error bars in figures (b), (c) and (d)
represent SD 
of the optimal energy levels obtained in the present work.
Horizontal solid lines represent zero level, and horizontal dash-dotted lines
--- error estimate $0.05$ cm$^{-1}$ made in \cite{Crosswhite}.
$N$ --- rotational quantum number (for the lower levels in figure (a)).
The combination differences $\Delta_2''(N)$ for some levels are also 
shown in figures (b) and (c).}\label{de_crosswhite}
\end{figure}

Sometimes the nature of the differences $\Delta E$  may be easily 
recognized. One typical example is illustrated in 
figure~\ref{de_crosswhite}, in which the differences $\Delta E$ are 
shown for rotational levels in the ground 
vibrational states ($v = 0$) of the $a^3\Sigma_g^+$ (d), $c^3\Pi_u^-$ (c), 
and $j^3\Delta_u^-$ (b) electronic states. The error bars represent our
estimates (one SD) of the empirical determination of the optimal energy
level values obtained by pure statistical approach in the present work. 
One may see that, as it was already mentioned,
quite often the deviations $\Delta E$ are less than $0.05$ cm$^{-1}$, but 
significantly higher than SD uncertainties of optimal level values
(figure~\ref{de_crosswhite}(d)).
In figure~\ref{de_crosswhite}(b) and figure~\ref{de_crosswhite}(c) one may see 
interesting behaviour of the deviation $\Delta E$ --- periodical changes
for rotational level with odd and even values of $N$
(for uncoupled levels of ortho- and para- molecules).
Similar effect is observed for many other electronic-vibrational 
states including the case shown in figure~\ref{de_crosswhite}(d). 

Small alternations of the deviations $\Delta E$  
(less than $0.1$ cm$^{-1}$) are connected
with the difference in the values of the shift between levels
of orto- and para- molecules obtained in the present work and
in \cite{Crosswhite}. More pronounced cases of the
alternations appear as a result of experimental errors due to
non-optimal method of energy level derivation used in 
\cite{Crosswhite} --- the sequential, multi-step procedure based on
calculating common differences. The error appeared in one common
difference due to one spurious experimental wavenumber value
is transferred from previous energy level value to the next one.
The mechanism is valid also for small random errors with various
final results, because subsequent pairs of lines within a band may 
have experimental errors of various values and sings.
Sometimes random errors may partially 
compensate each other, but in principle 
this method has the possibility of an accumulation of experimental 
wavenumber errors and transferring them into the errors of 
the empirical values of rovibronic energy levels. 

The typical example of the alternation appearance caused by 
spurious experimental wavenumber values is illustrated
in figure~\ref{de_crosswhite}(a,b,c). They demonstrate the influence of
two significant errors in wavenumbers of the $R$-branch lines
of the ($0-0$) band of the $j^3\Delta_g^- - c^3\Pi_u^-$
electronic transition on the empirical values of the
rotational levels obtained in \cite{Crosswhite}
for the upper $j^3\Delta_g^-$, $v = 0$ and lower
$c^3\Pi_u^-$, $0$ electronic-vibrational states of the band.

Excellent agreement between the observed and calculated 
wavenumbers reported in \cite{Crosswhite} seen in
figure~\ref{de_crosswhite}(a) is the evidence that the 
wavenumbers of the $R4$ and $R12$ spectral lines were
used in determination of the level values. 
Experimental wavenumbers reported in \cite{Crosswhite} for both
lines were classified as outliers by our statistical analysis.
Results of our independent experiments confirmed that this
is caused by occasional errors in \cite{Crosswhite}. The wavenumber 
values from \cite{Crosswhite} for the lines $R4$ and $R12$ are
underestimated and overestimated, respectively
(see figure~\ref{de_crosswhite}(a) and table~\ref{NewLines}). 
From the figure~\ref{de_crosswhite}(b,c) one may see how those 
experimental
errors are transferred into the wrong values of 
the empirical energy levels
via sequential adding of the common differences $\Delta_2'(N)$ 
and $\Delta_2''(N)$ calculated 
for upper and lower rovibronic levels. The cases of appearance, 
accumulation
and partial compensation of experimental errors may be found in 
figure~\ref{de_crosswhite}(b,c).

\begin{figure}[ht!]
\begin{center}
\epsfig{file=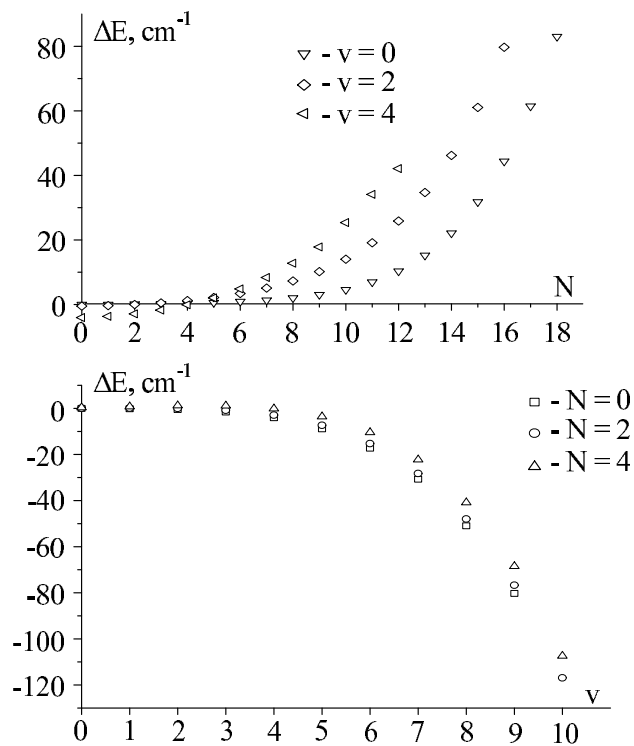, width=0.5\columnwidth,clip}
\end{center}
\caption{ The differences $\Delta E$ between optimal values of the energy 
levels obtained in the present work and those calculated with 
molecular constants from \cite{Nist} for various rotational and vibrational 
levels of the $a^3\Sigma_g^+$ electronic state of $D_2$ molecule.} \label{de_molconst}
\end{figure}

Taking into account the results of our previous studies of 
some triplet states of $H_2$ molecule \cite{LPU1986} it is 
easy to expect that the Dunham series 
expansions with empirical
molecular constants from \cite{Huber, Nist} have rather 
limited ability to describe rovibronic levels of
the $D_2$ isotopomer. These expectations were confirmed
by the results of the present work. Typical example of
the contradictions is illustrated in figure~\ref{de_molconst}
representing the differences $\Delta E$ between optimal 
values of the energy 
levels obtained in the present work and those calculated with 
molecular constants from \cite{Huber, Nist} for various 
rotational and vibrational 
levels of the $a^3\Sigma_g^+$ electronic state of $D_2$ molecule.
One may see that the deviations show strong dependencies
on vibrational and 
rotational quantum numbers, achieving values of about
100 cm$^{-1}$. It should be stressed that according to our
previous studies of the potential curves \cite{DLPU1987}
the $a^3\Sigma_g^+$  electronic state
is almost free from non-adiabatic effect of perturbations.
Higher electronic states are mainly perturbed due to 
electronic-rotational and electronic-vibrational interactions,
and simple Dunham expansions are non applicable. 

For determination of precise
absolute values of the triplet rovibronic
levels of $D_2 $ it is desirable to make pure statistical analysis
of all currently available experimental data on the wavenumbers of
singlet rovibronic transitions and to use the data of anticrossing 
spectroscopy \cite{JLDFMZ1976, MFZ1976, MZF1978} for establishing
the link between singlet and triplet state. The approach used in
the present work is most appropriate for solving this problem
in the future. Nowadays the absolute values of the triplet rovibronic 
levels (relative to the $X^1 \Sigma_g + $, $v = 0$
$N = 0$ ground rovibronic state) may be obtained only
by adding an absolute value $E_{a00}$ of 
energy of the $a^3\Sigma_g^+$, $v = 0$ $N = 0$
rovibronic level to the relative energy level values
obtained in the present work. According to \cite {Crosswhite}
$E_{a00}$ = $95348.2(4)$ cm$^{-1}$, while in 
\cite{JDHV1990} $E_{a00}$ = $95348.3(1)$ cm$^{-1}$.
These values coincide within the uncertainties reported in
original papers and may be used. The precision of such 
absolute calibration should be better than $1$ cm$^{-1}$.

Our relative values of triplet rovibronic levels of $D_2$ molecule 
may be used not only in comparisons with results of 
non-empirical calculations.
They are recommended for accurate calculating the wavenumbers of 
triplet-triplet optical transitions for computer simulations of 
emission and absorption spectra in applied spectroscopy of 
non-equilibrium gases and plasmas. Moreover, the high precision of
optimal level values achieved in the present work provides the opportunity
to expand existing identification of triplet rovibronic lines by
accurate calculating wavenumbers of all currently unassigned
rovibronic transitions allowed by the selection rules 
for electric dipole transitions
and detecting corresponding lines in experimental 
spectra. From the Grotrian diagram of the $D_2$ molecule it may be seen
that several new band systems may be discovered and investigated.

\section*{Acknowledgments}

The authors are indebted to Prof. S. Ross for providing  
the electronic version of the wavelength tables for lines of 
the $D_2$ emission spectrum measured and assigned by G.H. Dieke 
and co-workers (Appendix C. from Ref.\cite{Crosswhite}), and to 
Mr. M.S. Ryazanov for development of the computer code and helpful 
discussions. This work was supported, in part, by the Russian 
Foundation for Basic Research (the Grant \#06-03-32663a). 

\thispagestyle{empty}

\newpage

\newpage

\newpage

\newpage

\newpage

\newpage

\newpage

\newpage

\newpage

\newpage


\begin{table}
\caption{\label{NewLines}
Experimental values of the wavenumbers (in cm$^{-1}$) of some spectral lines of
$D_2$ obtained in the present work (P.W.)and those reported in \cite{Crosswhite}.
O-C denotes the differences between Observed wavenumber values and those Calculated
as differences of corresponding optimal energy level values obtained in the present work.}

\begin{tabular}{@{}lllrrrr}
\hline
 Electronic transition        &  Band    &  Line   &  \cite{Crosswhite}  &  O-C    &  P.W.              &  O-C    \\
\hline
$e^3\Sigma_u^+-a^3\Sigma_g^+$ & 2-0      & P11     & 13751.08            & -0.14   & 13751.29           &  0.07   \\
                              & 3-0      & R6      & 15868.98            & -0.05   & 15869.01           & -0.02   \\
                              &          & R9      & 15705.21            &  0.23   & 15704.98           &  0.00   \\
                              & 4-1      & P8      & 14862.89            & -0.08   & 14862.95           & -0.02   \\
                              & 5-2      & P6      & 14524.31            &  0.07   & 14524.26           &  0.02   \\	
                              & 6-2      & R4      & 15977.48            & -0.06   & 15977.56           &  0.02   \\
                              & 6-3      & P9      & 13703.95            &  0.09   & 13703.87           &  0.01   \\
                              & 7-3      & P5      & 15107.67            & -0.04   & 15107.71           &  0.00   \\
                              & 8-3      & R0      & 16354.21            &  0.05   & 16354.19           &  0.03   \\
                              & 8-4      & P5      & 14468.77            & -0.10   & 14468.86           & -0.01   \\
                              &          & R3      & 14727.33            & -0.06   & 14727.38           & -0.01   \\
                              & 9-5      & P4      & 13861.27            & -0.19   & 13861.45           & -0.01   \\
                              &          & R4      & 14002.82            & -0.11   & 14002.91           & -0.02   \\
                              & 10-4     & P4      & 16124.14            &  0.25   & 16123.94           &  0.05   \\
$f^3\Sigma_u^+-a^3\Sigma_g^+$ & 1-1      & R2      & 20346.49            &  0.07   & 20346.42           &  0.00   \\
                              & 1-2      & R0      & 18564.54            &  0.21   & 18564.19           & -0.14   \\
                              & 2-1      & P3      & 21653.13            &  0.20   & 21653.04           &  0.11   \\
                              & 2-3      & P3      & 18242.42            &  0.09   & 18242.40           &  0.07   \\
$d^3\Pi_u-a^3\Sigma_g^+$      & 0-1      & R9      & 15019.54            &  0.10   & 15019.44           &  0.00   \\
                              & 1-0      & P2      & 18207.70            &  0.10   & 18207.60           &  0.00   \\
                              &          & P5      & 18065.38            &  0.06   & 18065.36           &  0.04   \\
                              & 1-1      & Q1      & 16460.87            & -0.07   & 16460.93           & -0.01   \\
                              &          & Q10     & 16267.60            & -0.05   & 16267.64           & -0.01   \\
                              & 1-2      & Q10     & 14582.86            &  0.03   & 14582.84           &  0.01   \\
                              & 2-0      & Q1      & 19823.53            &  0.09   & 19823.45           &  0.01   \\
                              & 2-1      & R1      & 18065.70            &  0.07   & 18065.68           &  0.05   \\
                              &          & R5      & 18109.66            &  0.04   & 18109.64           &  0.02   \\
                              &          & R6      & 18107.89            & -0.04   & 18107.97           &  0.04   \\
                              &          & R8      & 18089.06            & -0.05   & 18089.10           & -0.01   \\
                              & 2-2      & R7      & 16386.78            &  0.05   & 16386.78           &  0.05   \\
                              &          & R8      & 16383.63            &  0.05   & 16383.58           &  0.00   \\
                              & 2-3      & Q6      & 14538.55            &  0.05   & 14538.51           &  0.01   \\
                              &          & Q9      & 14481.26            &  0.09   & 14481.19           &  0.02   \\
                              & 3-1      & R4      & 19576.76            &  0.17   & 19576.61           &  0.02   \\
                              & 3-2      & R4      & 17842.03            & -0.08   & 17842.10           & -0.01   \\
                              & 3-3      & R5      & 16180.93            & -0.10   & 16181.08           &  0.05   \\
                              & 3-4      & Q6      & 14415.64            &  0.15   & 14415.48           & -0.01   \\
                              & 4-2      & P8      & 18788.84            & -0.05   & 18788.84           & -0.05   \\
\hline
\end{tabular}
\end{table}

\begin{table}

\begin{tabular}{@{}lllrrrr}
\multicolumn{7}{l}{\bf table~\ref{NewLines}. \rm (Continued.)} \\
\hline
 Electronic transition        &  Band    &  Line   &  \cite{Crosswhite}  &  O-C    &  This work         &  O-C    \\
\hline
$d^3\Pi_u-a^3\Sigma_g^+$      & 4-3      & P3      & 17401.87            &  0.11   & 17401.74           & -0.02   \\
                              &          & R6      & 17586.74            &  0.07   & 17586.69           &  0.02   \\
                              & 4-5      & Q7      & 14280.99            &  0.06   & 14280.94           &  0.01   \\
                              & 5-2      & Q3      & 20510.44            & -0.14   & 20510.57           & -0.01   \\
                              & 5-3      & Q6      & 18757.31            & -0.17   & 18757.49           &  0.01   \\
                              &          & Q8      & 18677.56            &  0.08   & 18677.49           &  0.01   \\
                              & 6-4      & Q7      & 18420.01            & -0.11   & 18420.10           & -0.02   \\
                              & 7-6      & Q5      & 16727.50            &  0.18   & 16727.38           &  0.06   \\
$k^3\Pi_u-a^3\Sigma_g^+$      & 0-0      & P8      & 21966.23            & -0.01   & 21966.21           & -0.03   \\
                              &          & Q8      & 22180.52            &  0.16   & 22180.39           &  0.03   \\
                              &          & R6      & 22452.89            &  0.17   & 22452.75           &  0.03   \\
                              & 0-1      & Q8      & 20406.67            & -0.04   & 20406.68           & -0.03   \\
                              &          & R3      & 20613.91            &  0.19   & 20613.78           &  0.06   \\
                              & 1-2      & P4      & 20217.44            &  0.11   & 20217.32           & -0.01   \\
                              & 2-3      & P2      & 20142.72            & -0.14   & 20142.76           & -0.10   \\
                              &          & R2      & 20282.53            &  0.21   & 20282.45           &  0.13   \\
                              &          & R4      & 20320.54            &  0.09   & 20320.48           &  0.03   \\
                              & 3-3      & Q6      & 21592.35            &  0.22   & 21592.11           & -0.02   \\
                              & 4-3      & R4      & 23156.99            &  0.10   & 23156.92           &  0.03   \\
                              & 4-5      & P5      & 19755.38            & -0.15   & 19755.42           & -0.11   \\
                              & 5-4      & Q4      & 22761.98            & -0.13   & 22762.13           &  0.02   \\
$n^3\Pi_u-a^3\Sigma_g^+$      & 1-2      & Q5      & 22881.68            & -0.05   & 22881.71           & -0.02   \\
                              & 2-3      & P3      & 22670.82            & -0.12   & 22670.88           & -0.06   \\
$h^3\Sigma_g^+-c^3\Pi_u$      & 3-3      & P4      & 16779.23            & -0.09   & 16779.32           &  0.00   \\
$g^3\Sigma_g^+-c^3\Pi_u$      & 0-0      & P5      & 16588.93            &  0.13   & 16588.87           &  0.07   \\
                              & 0-1      & P3      & 15064.96            & -0.18   & 15065.27           &  0.13   \\
                              & 1-1      & P4      & 16502.17            &  0.13   & 16502.08           &  0.04   \\
                              & 2-1      & P5      & 17846.80            & -0.27   & 17847.07           &  0.00   \\
                              & 2-2      & P6      & 16164.95            &  0.22   & 16164.69           & -0.04   \\
                              &          & R9      & 16314.06            &  0.01   & 16314.07           &  0.02   \\
                              &          & Q10     & 16045.63            & -0.03   & 16045.64           & -0.02   \\
                              & 3-2      & P4      & 17652.53            &  0.09   & 17652.43           & -0.01   \\
                              &          & Q2      & 17801.97            & -0.06   & 17801.97           & -0.06   \\
                              &          & Q3      & 17765.57            & -0.05   & 17765.67           &  0.05   \\
                              & 3-3      & Q7      & 16024.02            &  0.09   & 16023.98           &  0.05   \\
                              &          & R2      & 16293.82            & -0.08   & 16293.88           & -0.02   \\
                              &          & R6      & 16210.71            & -0.04   & 16210.70           & -0.05   \\
$i^3\Pi_g^+-c^3\Pi_u$         & 0-0      & P11     & 16902.38            & -0.03   & 16902.34           & -0.07   \\
                              &          & Q10     & 17214.20            &  0.13   & 17214.14           &  0.07   \\
                              & 1-0      & P7      & 18524.43            &  0.16   & 18524.27           &  0.00   \\
                              &          & Q5      & 18728.49            &  0.20   & 18728.35           &  0.06   \\
                              &          & R1      & 18750.25            & -0.20   & 18750.51           &  0.06   \\
                              &          & R7      & 18957.79            &  0.17   & 18957.66           &  0.04   \\
                              & 2-1      & Q2      & 18425.10            & -0.17   & 18425.19           & -0.08   \\
                              & 3-3      & P8      & 16450.63            & -0.08   & 16450.62           & -0.09   \\
                              &          & R6      & 16849.79            &  0.10   & 16849.78           &  0.09   \\
\hline
\end{tabular}
\end{table}

\begin{table}

\begin{tabular}{@{}lllrrrr}
\multicolumn{7}{l}{\bf table~\ref{NewLines}. \rm (Continued.)} \\
\hline
 Electronic transition        &  Band    &  Line   &  \cite{Crosswhite}  &  O-C    &  This work         &  O-C    \\
\hline
$i^3\Pi_g^--c^3\Pi_u$         & 0-0      & P7      & 16787.90            & -0.10   & 16787.97           & -0.03   \\
                              &          & P14     & 16369.80            & -0.08   & 16369.84           & -0.04   \\
                              &          & Q3      & 17071.29            & -0.09   & 17071.28           & -0.10   \\
                              &          & Q9      & 16898.78            &  0.13   & 16898.69           &  0.04   \\
                              &          & R9      & 17150.62            &  0.17   & 17150.53           &  0.08   \\
                              &          & R12     & 17109.51            & -0.09   & 17109.64           &  0.04   \\
                              & 1-0      & P3      & 18525.88            &  0.22   & 18525.72           &  0.06   \\
                              &          & Q3      & 18598.23            & -0.27   & 18598.59           &  0.09   \\
                              & 2-1      & R1      & 18451.87            &  0.23   & 18451.64           &  0.00   \\
                              &          & R2      & 18464.46            &  0.13   & 18464.34           &  0.01   \\
                              & 2-2      & P10     & 16345.66            & -0.04   & 16345.72           &  0.02   \\
                              &          & R4      & 16864.08            &  0.21   & 16863.92           &  0.05   \\
                              &          & R8      & 16857.89            & -0.59   & 16858.46           & -0.02   \\
                              & 2-3      & Q3      & 15207.72            &  0.05   & 15207.75           &  0.08   \\
                              & 3-3      & P3      & 16517.58            &  0.06   & 16517.57           &  0.05   \\
                              &          & Q2      & 16599.70            &  0.05   & 16599.70           &  0.05   \\
                              &          & R5      & 16693.01            & -0.16   & 16693.13           & -0.04   \\
$j^3\Delta_g^+-c^3\Pi_u$      & 0-0      & P13     & 17309.93            & -0.02   & 17309.93           & -0.02   \\
                              &          & Q12     & 17667.66            &  0.00   & 17667.68           &  0.02   \\
                              & 1-0      & Q2      & 19060.17            &  0.14   & 19060.15           &  0.12   \\
                              & 1-1      & P10     & 17192.24            &  0.02   & 17192.26           &  0.04   \\
                              &          & Q7      & 17439.67            &  0.11   & 17439.65           &  0.09   \\
                              &          & R8      & 17722.82            & -0.04   & 17722.82           & -0.04   \\
                              & 2-1      & Q5      & 18907.85            & -0.11   & 18907.90           & -0.06   \\
                              & 2-2      & P9      & 17081.06            & -0.10   & 17081.14           & -0.02   \\
$j^3\Delta_g^--c^3\Pi_u$      & 0-0      & P14     & 17182.21            & -0.01   & 17182.26           &  0.04   \\
                              &          & Q13     & 17564.20            & -0.02   & 17564.24           &  0.02   \\
                              &          & R4      & 17654.95            & -0.13   & 17655.06           & -0.02   \\
                              &          & R12     & 17922.15            &  0.22   & 17921.87           & -0.06   \\
                              & 1-0      & P7      & 18863.60            &  0.26   & 18863.37           &  0.03   \\	
                              &          & Q4      & 19066.14            & -0.18   & 19066.26           & -0.06   \\
                              & 1-1      & Q8      & 17423.59            &  0.10   & 17423.54           &  0.05   \\
                              &          & Q10     & 17431.73            & -0.09   & 17431.78           & -0.04   \\
                              &          & R9      & 17709.08            & -0.46   & 17709.58           &  0.04   \\
                              & 2-1      & P4      & 18786.75            & -0.14   & 18786.94           &  0.05   \\
                              &          & Q4      & 18903.82            & -0.28   & 18904.14           &  0.04   \\
                              &          & Q5      & 18903.31            & -0.22   & 18903.52           & -0.01   \\
                              & 2-2      & P3      & 17200.68            & -0.10   & 17200.72           & -0.06   \\
                              & 2-3      & R1      & 15783.87            & -0.25   & 15784.09           & -0.03   \\
                              & 3-3      & P3      & 17107.09            & -0.04   & 17107.11           & -0.02   \\
                              & 3-4      & Q2      & 15692.04            &  0.16   & 15691.96           &  0.08   \\
\hline
\end{tabular}
\end{table}


\newpage

\begin{table}
\caption{\label{LineNumber}
The amounts of various kinds of the experimental data reported in 
various original papers and obtained in the present work (P.W.): 
total \it reported\rm,
\it useable \rm for the statistical analysis, \it excluded \rm 
as the outliers, \it used \rm for determination of optimal level 
values, and the rate of used lines in $\%$.}

\begin{tabular}{@{}lrrrrrrrrrr}
\hline
Reference & \cite{Crosswhite} & \cite{Dieke} & \cite{DiekeBlue} & \cite{DabrHerz} & 
            \cite{DiekePorto} & \cite{GloersenDieke} & \cite{Davies} & \cite{FreundMiller} & 
            P.W. & Total \\
\hline
reported  & 3117 &  350 &  285 &   81 &   37 &   83 &   3 &   1 &  125 & 4082 \\
useable   & 3053 &  350 &  285 &   81 &   37 &   12 &   3 &   1 &  125 & 3947 \\
excluded  &  179 &   18 &   18 &    5 &    4 &   10 &   0 &   0 &    0 &  234 \\
used      & 2874 &  332 &  267 &   76 &   33 &    2 &   3 &   1 &  125 & 3713 \\
$\%$      & 94.1 & 94.9 & 93.7 & 93.8 & 89.2 & 16.7 & 100 & 100 &  100 & 94.1 \\
\hline
\end{tabular}
\end{table}


\newpage

\begin{table}															
\caption{\label{NewLevels}
Optimal values $E_{nvN}$ of rovibronic energy levels (in cm$^{-1}$)
for various triplet electronic states of $D_2$ molecule obtained in the present work.
The uncertainties of the $E_{nvN}$ value determination (one SD) are shown in brackets
in units of last significant digit. $n_{\nu}$ --- the number of various spectral lines 
directly connected with certain level. $\Delta E$ is the difference
between energy level values obtained in the present work and those
reported in \cite{Crosswhite}.}
                                                                                                                                
\begin{tabular}{@{}rr@{}c@{}lrrr@{}c@{}lrrr@{}c@{}lrr}
\hline                                                                                                                             
\multicolumn{16}{c}{$(1s\sigma 2s\sigma)a^3\Sigma_g^+$} \\
\hline
 & \multicolumn{5}{c}{v = 0} & \multicolumn{5}{c}{v = 1} & \multicolumn{5}{c}{v = 2} \\
\hline
 N & \multicolumn{3}{c}{ $E_{nvN}$ }& $n_{\nu}$ & $\Delta E$ &
     \multicolumn{3}{c}{ $E_{nvN}$ }& $n_{\nu}$ & $\Delta E$ &
     \multicolumn{3}{c}{ $E_{nvN}$ }& $n_{\nu}$ & $\Delta E$ \\
\hline
 0 &    0 & &              & 21  &  0     &  1814 &.& 908(5)  & 25    & -0.008 & 3560 &.& 726(6)  & 26    & -0.016   \\
 1 &   33 &.& 631(4)       & 43  & -0.011 &  1847 &.& 364(8)  & 55    & -0.014 & 3592 &.& 036(8)  & 56    & -0.016   \\
 2 &  100 &.& 770(6)       & 52  & -0.010 &  1912 &.& 172(5)  & 66    & -0.012 & 3654 &.& 573(5)  & 70    & -0.003   \\
 3 &  201 &.& 109(9)       & 47  & -0.009 &  2009 &.& 042(8)  & 63    & -0.002 & 3748 &.& 034(8)  & 64    & -0.004   \\
 4 &  334 &.& 333(6)       & 49  & -0.013 &  2137 &.& 628(6)  & 57    & -0.008 & 3872 &.& 103(6)  & 64    & -0.003   \\
\hline
\end{tabular}
\end{table}

\begin{table}

\begin{tabular}{@{}rr@{}c@{}lrrr@{}c@{}lrrr@{}c@{}lrr}
\multicolumn{16}{l}{\bf table~\ref{NewLevels}. \rm (Continued.)} \\
\hline                                                                                                                             
\multicolumn{16}{c}{$(1s\sigma 2s\sigma)a^3\Sigma_g^+$} \\
\hline
 & \multicolumn{5}{c}{v = 6} & \multicolumn{5}{c}{v = 7} & \multicolumn{5}{c}{v = 8} \\
\hline
 N & \multicolumn{3}{c}{ $E_{nvN}$ }& $n_{\nu}$ & $\Delta E$ &
     \multicolumn{3}{c}{ $E_{nvN}$ }& $n_{\nu}$ & $\Delta E$ &
     \multicolumn{3}{c}{ $E_{nvN}$ }& $n_{\nu}$ & $\Delta E$  \\
\hline
 0 &  9880 &.& 806(11) & 10         & -0.016 & 11299 &.& 845(16) & 6       & -0.045 & 12655 &.& 467(18) & 5       & -0.007 \\
 1 &  9907 &.& 775(10) & 26         &  0.005 & 11325 &.& 784(11) & 18      & -0.004 & 12680 &.& 356(12) & 12      & -0.036 \\
 2 &  9961 &.& 617(7)  & 32         & -0.017 & 11377 &.& 500(8)  & 18      &  0.000 & 12729 &.& 972(9)  & 13      & -0.052 \\
 3 & 10042 &.& 093(10) & 28         & -0.013 & 11454 &.& 821(12) & 14      & -0.010 & 12804 &.& 129(13) & 14      & -0.059 \\
\hline
\end{tabular}
\end{table}

\begin{table}

\begin{tabular}{@{}rr@{}c@{}lrrr@{}c@{}lrrr@{}c@{}lrr}
\multicolumn{16}{l}{\bf table~\ref{NewLevels}. \rm (Continued.)} \\
\hline                                                                                                                             
\multicolumn{16}{c}{$(1s\sigma 2p\pi)c^3\Pi_u^-$} \\ 
\hline
 & \multicolumn{5}{c}{v = 3} & \multicolumn{5}{c}{v = 4} & \multicolumn{5}{c}{v = 5} \\ 
\hline
 N & \multicolumn{3}{c}{ $E_{nvN}$ }& $n_{\nu}$ & $\Delta E$ &
     \multicolumn{3}{c}{ $E_{nvN}$ }& $n_{\nu}$ & $\Delta E$ &
     \multicolumn{3}{c}{ $E_{nvN}$ }& $n_{\nu}$ & $\Delta E$  \\ 
\hline
 1 & 4727 &.& 022(9)  & 7             &  0.108 &  6226 &.& 366(19) & 3              &  0.144 & 7667 &.& 225(13) & 3            &  0.125 \\
 2 & 4782 &.& 112(9)  & 10            &  0.038 &  6279 &.& 454(11) & 7              &  0.096 & 7718 &.& 429(10) & 2            &  0.141 \\
 3 & 4864 &.& 307(13) & 11            &  0.073 &  6358 &.& 796(14) & 4              &  0.084 & 7794 &.& 944(13) & 2            &  0.116 \\
 4 & 4973 &.& 456(10) & 12            &  0.054 &  6464 &.& 117(13) & 4              &  0.063 & 7896 &.& 475(10) & 2            & -0.035 \\
\hline
\end{tabular}
\end{table}

\begin{table}

\begin{tabular}{@{}rr@{}c@{}lrrr@{}c@{}lrrr@{}c@{}lrr}
\multicolumn{16}{l}{\bf table~\ref{NewLevels}. \rm (Continued.)} \\
\hline                                                                                                                             
\multicolumn{16}{c}{$(1s\sigma 2p\pi)c^3\Pi_u^+$} \\ 
\hline
 & \multicolumn{5}{c}{v = 3} & \multicolumn{5}{c}{v = 4} & \multicolumn{5}{c}{v = 5} \\ \hline
 N & \multicolumn{3}{c}{ $E_{nvN}$ }& $n_{\nu}$ & $\Delta E$ &
     \multicolumn{3}{c}{ $E_{nvN}$ }& $n_{\nu}$ & $\Delta E$ &
     \multicolumn{3}{c}{ $E_{nvN}$ }& $n_{\nu}$ & $\Delta E$  \\ 
\hline
 1 & 4727 &.& 078(7)  & 14             &  0.052 & 6226 &.& 429(10) & 7              &  0.081 & 7667 &.& 297(13) & 2             &  0.053 \\
 2 & 4782 &.& 081(12) & 16             &  0.109 & 6279 &.& 553(10) & 9              &  0.007 & 7718 &.& 544(13) & 2             &  0.016 \\
 3 & 4864 &.& 446(8)  & 15             &  0.014 & 6359 &.& 007(11) & 7              &  0.133 & 7795 &.& 177(10) & 1             &  -     \\
 4 & 4973 &.& 606(14) & 9              &  0.084 & 6464 &.& 401(14) & 8              &  0.079 & 7896 &.& 833(13) & 1             &  -     \\
\hline
\end{tabular}
\end{table}

\begin{table}

\begin{tabular}{@{}rr@{}c@{}lrrr@{}c@{}lrrr@{}c@{}lrr}
\multicolumn{16}{l}{\bf table~\ref{NewLevels}. \rm (Continued.)} \\
\hline                                                                                                                             

\multicolumn{16}{c}{$(1s\sigma 3p\sigma)e^3\Sigma_u^+$} \\ 
\hline
 & \multicolumn{5}{c}{v = 0} & \multicolumn{5}{c}{v = 1} & \multicolumn{5}{c}{v = 2} \\ 
\hline
 N & \multicolumn{3}{c}{ $E_{nvN}$ }& $n_{\nu}$ & $\Delta E$ &
     \multicolumn{3}{c}{ $E_{nvN}$ }& $n_{\nu}$ & $\Delta E$ &
     \multicolumn{3}{c}{ $E_{nvN}$ }& $n_{\nu}$ & $\Delta E$  \\ 
\hline
 0 & 11649 &.& 190(14) & 5            & -0.010 & 13137 &.& 502(10) & 6       & -0.012 & 14558 &.& 258(12) & 6    &  0.002  \\
 1 & 11676 &.& 174(8)  & 12           & -0.004 & 13163 &.& 431(6)  & 12      & -0.011 & 14583 &.& 137(7)  & 14   &  0.003  \\
 2 & 11730 &.& 037(13) & 12           & -0.007 & 13215 &.& 170(9)  & 15      & -0.010 & 14632 &.& 770(11) & 13   &  0.000  \\
 3 & 11810 &.& 600(9)  & 17           & -0.020 & 13292 &.& 570(7)  & 16      & -0.010 & 14707 &.& 003(8)  & 15   & -0.003  \\
\hline
\end{tabular}
\end{table}

\begin{table}

\begin{tabular}{@{}rr@{}c@{}lrrr@{}c@{}lrrr@{}c@{}lrr}
\multicolumn{16}{l}{\bf table~\ref{NewLevels}. \rm (Continued.)} \\
\hline                                                                                                                             

\multicolumn{16}{c}{$(1s\sigma 3p\sigma)e^3\Sigma_u^+$} \\ 
\hline
 & \multicolumn{5}{c}{v = 6} & \multicolumn{5}{c}{v = 7} & \multicolumn{5}{c}{v = 8} \\ 
\hline
 N & \multicolumn{3}{c}{ $E_{nvN}$ }& $n_{\nu}$ & $\Delta E$ &
     \multicolumn{3}{c}{ $E_{nvN}$ }& $n_{\nu}$ & $\Delta E$ &
     \multicolumn{3}{c}{ $E_{nvN}$ }& $n_{\nu}$ & $\Delta E$  \\ 
\hline
 0 & 19546 &.& 641(11) & 7        & -0.001 & 20605 &.& 338(10) & 8         & -0.008 & 21575 &.& 310(13) & 9      & -0.030  \\
 1 & 19567 &.& 047(7)  & 16       & -0.007 & 20624 &.& 515(6)  & 15        & -0.015 & 21593 &.& 153(7)  & 19     & -0.013  \\
 2 & 19607 &.& 728(10) & 16       & -0.008 & 20662 &.& 787(9)  & 16        & -0.017 & 21628 &.& 745(10) & 19     & -0.005  \\
 3 & 19668 &.& 569(7)  & 16       &  0.001 & 20720 &.& 289(6)  & 14        & -0.009 & 21681 &.& 957(7)  & 20     & -0.007  \\
\hline
\end{tabular}
\end{table}

\begin{table}

\begin{tabular}{@{}rr@{}c@{}lrrr@{}c@{}lrrr@{}c@{}lrr}
\multicolumn{16}{l}{\bf table~\ref{NewLevels}. \rm (Continued.)} \\
\hline                                                                                                                             

\multicolumn{16}{c}{$(1s\sigma 3p\pi)d^3\Pi_u^-$} \\ 
\hline
 & \multicolumn{5}{c}{v = 3} & \multicolumn{5}{c}{v = 4} & \multicolumn{5}{c}{v = 5} \\ \hline
 N & \multicolumn{3}{c}{ $E_{nvN}$ }& $n_{\nu}$ & $\Delta E$ &
     \multicolumn{3}{c}{ $E_{nvN}$ }& $n_{\nu}$ & $\Delta E$ &
     \multicolumn{3}{c}{ $E_{nvN}$ }& $n_{\nu}$ & $\Delta E$  \\ 
\hline
 1 & 21343 &.& 645(12) & 9       &  0.005 & 22769 &.& 442(12) & 8        & -0.012 & 24135 &.& 655(10) & 10    &  0.005  \\
 2 & 21396 &.& 902(9)  & 8       & -0.002 & 22820 &.& 678(8)  & 8        &  0.002 & 24184 &.& 948(7)  & 10    & -0.008  \\
 3 & 21476 &.& 477(12) & 8       & -0.007 & 22897 &.& 296(12) & 7        & -0.006 & 24258 &.& 610(10) & 9     &  0.000  \\
 4 & 21582 &.& 138(9)  & 8       & -0.008 & 22998 &.& 963(8)  & 8        & -0.013 & 24356 &.& 367(7)  & 9     & -0.017  \\
\hline
\end{tabular}
\end{table}

\begin{table}

\begin{tabular}{@{}rr@{}c@{}lrrr@{}c@{}lrrr@{}c@{}lrr}
\multicolumn{16}{l}{\bf table~\ref{NewLevels}. \rm (Continued.)} \\
\hline                                                                                                                             
\multicolumn{16}{c}{$(1s\sigma 3p\pi)d^3\Pi_u^+$} \\ 
\hline
 & \multicolumn{5}{c}{v = 0} & \multicolumn{5}{c}{v = 1} & \multicolumn{5}{c}{v = 2} \\ 
\hline
 N & \multicolumn{3}{c}{ $E_{nvN}$ }& $n_{\nu}$ & $\Delta E$ &
     \multicolumn{3}{c}{ $E_{nvN}$ }& $n_{\nu}$ & $\Delta E$ &
     \multicolumn{3}{c}{ $E_{nvN}$ }& $n_{\nu}$ & $\Delta E$  \\ 
\hline
 1 & 16695 &.& 884(8)  & 8            & -0.014 & 18308 &.& 374(7)  & 12         &  0.006 & 19857 &.& 301(6)  & 15           & -0.011 \\
 2 & 16755 &.& 611(11) & 8            & -0.011 & 18365 &.& 854(9)  & 14         & -0.004 & 19912 &.& 998(10) & 12           &  0.002 \\
 3 & 16844 &.& 912(9)  & 8            & -0.002 & 18451 &.& 709(7)  & 13         & -0.019 & 19996 &.& 263(6)  & 16           & -0.013 \\
 4 & 16963 &.& 424(13) & 5            & -0.004 & 18565 &.& 166(10) & 13         &  0.004 & 20106 &.& 670(9)  & 14           &  0.000 \\
\hline
\end{tabular}
\end{table}

\begin{table}

\begin{tabular}{@{}rr@{}c@{}lrrr@{}c@{}lrrr@{}c@{}lrr}
\multicolumn{16}{l}{\bf table~\ref{NewLevels}. \rm (Continued.)} \\
\hline                                                                                                                             

\multicolumn{16}{c}{$(1s\sigma 3d\sigma)g^3\Sigma_g^+$} \\ 
\hline
 & \multicolumn{5}{c}{v = 3} & \multicolumn{5}{c}{v = 4} \\ 
\hline
 N & \multicolumn{3}{c}{ $E_{nvN}$ }& $n_{\nu}$ & $\Delta E$ &
     \multicolumn{3}{c}{ $E_{nvN}$ }& $n_{\nu}$ & $\Delta E$  \\ 
\hline
 0 & 20982 &.& 138(17) & 1       &  0.052 &       & &         &        &         \\
 1 & 20996 &.& 599(14) & 3       &  0.091 &       & &         &        &         \\
 2 & 21027 &.& 241(9)  & 6       &  0.029 & 22276 &.& 495(14) & 3      &  0.085  \\
 3 & 21075 &.& 984(15) & 6       &  0.156 & 22327 &.& 149(19) & 2      &  0.001  \\
\hline
\end{tabular}
\end{table}

\begin{table}

\begin{tabular}{@{}rr@{}c@{}lrrr@{}c@{}lrrr@{}c@{}lrr}
\multicolumn{16}{l}{\bf table~\ref{NewLevels}. \rm (Continued.)} \\
\hline
\multicolumn{16}{c}{$(1s\sigma 3d\pi)i^3\Pi_g^+$} \\ 
\hline
 & \multicolumn{5}{c}{v = 0} & \multicolumn{5}{c}{v = 1} & \multicolumn{5}{c}{v = 2} \\ 
\hline
 N & \multicolumn{3}{c}{ $E_{nvN}$ }& $n_{\nu}$ & $\Delta E$ &
     \multicolumn{3}{c}{ $E_{nvN}$ }& $n_{\nu}$ & $\Delta E$ &
     \multicolumn{3}{c}{ $E_{nvN}$ }& $n_{\nu}$ & $\Delta E$  \\ 
\hline
 1 & 16998 &.& 00(2)   & 4       &  0.024 & 18539 &.& 50(2)   & 5            &  0.040 &       & &       &        &         \\
 2 & 17086 &.& 474(17) & 5       & -0.014 & 18617 &.& 862(19) & 6            &  0.008 & 20033 &.& 06(2) & 5      &  0.070  \\
 3 & 17202 &.& 627(19) & 5       & -0.007 & 18725 &.& 790(19) & 7            &  0.010 & 20130 &.& 70(3) & 3      &  0.080  \\
 4 & 17343 &.& 627(15) & 6       & -0.047 & 18860 &.& 377(15) & 7            &  0.033 & 20254 &.& 70(3) & 5      &  0.040  \\
\hline
\end{tabular}
\end{table}

\begin{table}

\begin{tabular}{@{}rr@{}c@{}lrrr@{}c@{}lrrr@{}c@{}lrr}
\multicolumn{16}{l}{\bf table~\ref{NewLevels}. \rm (Continued.)} \\
\hline                                                                                                                             

\multicolumn{16}{c}{$(1s\sigma 3d\delta)j^3\Delta_g^-$} \\ 
\hline
 & \multicolumn{5}{c}{v = 3} & \multicolumn{5}{c}{v = 4} & \multicolumn{5}{c}{v = 5} \\ 
\hline
 N & \multicolumn{3}{c}{ $E_{nvN}$ }& $n_{\nu}$ & $\Delta E$ &
     \multicolumn{3}{c}{ $E_{nvN}$ }& $n_{\nu}$ & $\Delta E$ &
     \multicolumn{3}{c}{ $E_{nvN}$ }& $n_{\nu}$ & $\Delta E$  \\ 
\hline
 2 & 21971 &.& 433(16) & 4       &  0.087 & 23370 &.& 57(3) & 3        &  0.070 & 24709 &.& 17(5)     & 2 &  0.070 \\
 3 & 22054 &.& 903(15) & 3       &  0.087 & 23448 &.& 57(3) & 3        &  0.090 &       & &           &   &        \\
 4 & 22165 &.& 210(16) & 4       &  0.080 & 23551 &.& 78(4) & 2        &  0.060 &       & &           &   &        \\
 5 & 22301 &.& 619(15) & 2       &  0.041 & 23679 &.& 34(4) & 2        &  0.030 &       & &           &   &        \\
\hline
\end{tabular}
\end{table}


\newpage

\begin{thebibliography}{99}

\bibitem{GLT1982}
Greben'kov V S, Lavrov B P and Tyutchev M V 1982
{\it Sov.J.Opt.Technol.} {\bf 49} 115--8

\bibitem{PBS2005}
Pospieszczyk A, Brezinsek S, Sergienko G {\it et al.} 2005
{\it J. of Nucl. Mat.} {\bf 337-339} 500--4

\bibitem{LKOR1997}
Lavrov B P, K\"aning M, Ovtchinnikov V L and R\"opcke J 1997
{\it International Workshop ''Frontiers in Low Temperature Plasma Diagnostics II''
Bad Honnef (Germany)} 
p~169--72

\bibitem{LMKR1999}
Lavrov B P, Melnikov A S, K\"aning M and R\"opcke J 1999
{\it Phys.Rev.E} {\bf 59} 3526--43

\bibitem{RDKL2001}
R\"opcke J, Davies P B, K\"aning M and Lavrov B P 2001
{\it Low Temperature Plasma Physics --- Fundamental Aspects and Applications
(Wiley-VCH, Berlin, New York, Toronto)} p 173--98.

\bibitem{LPR2006}
Lavrov B P, Pipa A V and R\"opcke J 2006
{\it Plasma Sources Science and Technology} {\bf 15} 135--46

\bibitem{Wetal2006}
Wyart J F, Meftah A, Bachelier A, Sinzelle J,
Tchang-Brillet W \"U L, Champion N, Spector N and Sugar J 2006
{\it J. Phys. B: At. Mol. Opt. Phys.} {\bf 39} L77--82

\bibitem{LavrovRiazanovJetf}
Lavrov B P and Ryazanov M R 2005
{\it JETP Letters} {\bf 81} 371--4

\bibitem{LavrovRiazanovOiS}
Lavrov B P and Ryazanov M R 2005
{\it Opt. Spektrosc.} {\bf 99} 890--6

\bibitem{Kovacs1969}
Kov\'acs I 1969
{\it Rotational structure in the spectra of diatomic molecules}
(London: Adam Hilger Ltd) p~320

\bibitem{Richardson}
Richardson O W 1934
{\it Molecular Hydrogen and Its Spectrum}
(Yale Univ. Press, New Haven)

\bibitem{Dieke1958}
Dieke G H 1958
{\it J. Mol. Spectrsc.} {\bf 2} 494--517

\bibitem{Dieke1972}
Dieke G H 1972
{\it The Hydrogen Molecule Wavelength Tables of
G.H.Dieke / Ed. by H.M.Crosswhite}
(N.Y.: Wiley) p~616

\bibitem{BredohlHerzberg1973}
Bredohl H and Herzberg G 1973
{\it Can. J. Phys.}  {\bf 51} 867--87

\bibitem{Huber}
Huber K P and Herzberg G 1979
{\it Molecular Spectra and Molecular Structure
V.~IV. Constants of Diatomic Molecules 
(Van Nostrand Reinhold Co, New York)} p~262--7

\bibitem{Crosswhite}
Freund R S, Schiavone J A and Crosswhite H M 1985
{\it J. Phys. Chem. Ref. Data.} {\bf 14} 235--383

\bibitem{RossJungenMatzkin2001}
Ross S C and Jungen Ch 2001
{\it Can. J. Phys.} {\bf 79} 561

\bibitem{LO1978}
Lavrov B P and Otorbaev J K 1978 
{\it Opt.Spectrosc.} {\bf 44} 360--1 

\bibitem{RLTBjcp2006}
Roudjane M, Launay F and Tchang-Brillet W-U L 2006
{\it J. Chem. Phys.} {\bf 125} 214305-1--9

\bibitem{RLTBjcp2007}
Roudjane M, Launay F and Tchang-Brillet W-U L 2007
{\it J. Chem. Phys.} {\bf 127} 054307-1--6

\bibitem{Azarov1991}
Azarov V I 1991
{\it Phys. Scr.} {\bf 44} 528--38

\bibitem{Azarov1993}
Azarov V I 1993
{\it Phys. Scr.} {\bf 48} 656--67

\bibitem{BHterms}
Lavrov B P and Ryazanov M S 2005
{\it physics/0504044 at http://arXiv.org}

\bibitem{ALMU2008}
Lavrov B P, Astashkevich S A, Modin A V and Umrikhin I S 2008
{\it Himicheskaya Fisika (Chem.Phys. ISSN 0207-401X, in Russian)} {\bf 27} 22--38

\bibitem{Nist}
{\it http://webbook.nist.gov/}

\bibitem{LPU1986}
Lavrov B P, Prosikhin V P and Ustimov V I 1986 
{\it Sov. Phys. J. (UK).} {\bf 29} 137-40

\bibitem{Albriton}
Albriton D L, Harrop W J, Schmeltekopf A L, Zare R N and Crow E L 1973
{\it J. Mol. Spectr.} {\bf 46} 67--88

\bibitem{DabrHerz}
Dabrowski I and Herzberg G 1984
{\it Acta Phys. Hung.} {\bf 55} 219--28

\bibitem{Davies}
Davies P B, Guest M A and Johnson S A 1987
{\it J. Chem. Phys.} {\bf 88} 2884--90

\bibitem{DLPU1987}
Drachev A I, Lavrov B P, Prosikhin V P and Ustimov V I 1987 
{\it Sov.J.Chem.Phys. (UK).} {\bf 4} 1663-75

\bibitem{DiekeBlue}
Dieke G H and Blue R W 1935
{\it Phys. Rev.} {\bf 47} 261--72

\bibitem{Dieke}
Dieke G H 1935
{\it Phys. Rev.} {\bf 48} 606--9

\bibitem{GloersenDieke}
Gloersen P and Dieke G H 1965
{\it J. Mol. Spectrosc.} {\bf 16} 191--204

\bibitem{DiekePorto}
Porto S P S and Dieke G H 1955
{\it J. Opt. Soc. Am.} {\bf 45} 447--50

\bibitem{FreundMiller}
Freund R S, Miller T A and Zegarski B R 1976
{\it J. Chem. Phys.} {\bf 64} 4069--75

\bibitem{Hudson}
Hudson D J 1964
{\it Statistics. Lectures on Elementary Statistics
and Probability}
(Geneva)

\bibitem{VanderSluis}
Vander Sluis K L 1966
{\it J. Opt. Soc. Am.} {\bf 56} 1600--03

\bibitem{RK1974}
Radziemski L J and Kaufman V 1974
{\it J. Opt. Soc. Am.} {\bf 64} 366--89

\bibitem{RA1965}
Radziemski L J and Andrew K L 1965
{\it J. Opt. Soc. Am.} {\bf 55} 474--91

\bibitem{RK1969}
Radziemski L J and Kaufman V 1969
{\it J. Opt. Soc. Am.} {\bf 59} 424--43

\bibitem{Aslund1965}
\AA slund N 1965
{\it Arkiv Fysik.} {\bf 30} 377--396

\bibitem{LosAlamos1970}
Radziemski L J, Fisher K J and Steinhaus D W 1970
{\it Calculation of Atomic-Energy-Level Values}
(Los Alamos National Laboratory Report No. LA-4402)

\bibitem{Wetal2007}
Wyart J F, Meftah A, Tchang-Brillet W \"U L,
Champion N, Lamrous O, Spector N and Sugar J 2007
{\it J. Phys. B: At. Mol. Opt. Phys.} {\bf 40} 3957--72

\bibitem{Aslund1974}
\AA slund N 1974
{\it J. Mol. Spectrosc.} {\bf 50} 424

\bibitem{Ross2003}
Ross S M 2003
{\it J. Engr. Technology} {\bf 20} Fall

\bibitem{LU2007}
Lavrov B P and Umrikhin I S 2007
{\it physics/07030114v2 at http://arXiv.org}

\bibitem{AKKKLOR1996}
Astashkevich S A, K\"aning M, K\"aning E, Kokina N V, Lavrov B P, Ohl A and R\"opcke J 1996
{\it JQSRT} {\bf 56} 725--51

\bibitem{LT1982}
Lavrov B P and Tyutchev M V 1982
{\it Sov. J. Opt. Technol.} {\bf 49} 741--43

\bibitem{JLDFMZ1976}
Jost R, Lombardi M, Derouard J, Freund R S, Miller T A and Zegarski B R 1976
{\it J. Chem. Phys. Lett.} {\bf 37} 507

\bibitem{MFZ1976}
Miller T A, Freund R S and Zegarski B R 1976
{\it J. Chem. Phys.} {\bf 64} 1842--7

\bibitem{MZF1978}
Miller T A, Zegarski B R and Freund R S 1978
{\it J. Mol. Spectrosc.} {\bf 69} 199

\bibitem{JDHV1990}
Jungen Ch, Dabrowski I, Herzberg G and Vervloet M 1990
{\it J. Chem. Phys.} {\bf 93} 2289--98

\end{thebibliography}
\end{document}